\documentstyle[11pt]{article}
\newcommand{\be}{\begin{equation}}
\newcommand{\ee}{\end{equation}}

\title{SOME REMARKS ON WHITHAM DYNAMICS}
\author{Robert Carroll\\University of Illinois\\ email:  rcarroll@math.uiuc.edu}

\date{April, 1998}

\begin{document}
\bibliographystyle{plain}
\maketitle
\begin{abstract} A number of formulas are displayed concerning Whitham
theory for a simple example of pure $N=2$ susy Yang-Mills (YM) with
gauge group $SU(2)$.  In particular this serves to illuminate the role
of $\Lambda$ and $T_1$ derivatives and the interaction with
prepotentials $F^{SW}$ (Seiberg-Witten) and $F^W$ (Whitham).
\end{abstract}

\section{INTRODUCTION}
\renewcommand{\theequation}{1.\arabic{equation}}\setcounter{equation}{0}

One knows that there are remarkable relations between integrable hierarchies,
$N=2$ susy YM, and Whitham theory (see e.g. \cite{bb,cc,cg,cb,co,
dc,ea,fb,gc,gf,ia,ka,kc,ko,lg,mb,na,ta}).  There is also a stringy
background involving Calabi-Yau (CY) manifolds, mirror symmetry,
and branes (see e.g. \cite{ge,kz,la}) but we go back to
more elementary matters involving SW curves in order to clarify the role
of Whitham dynamics (cf. also \cite{bc,By,bf,ba,Bz,be,dz,dg,
de,da,km,mc,ma,sd,se} for related papers). 
A basic reference is \cite{gf} which shows
how $\Lambda$ and $T_1$ derivatives are related in the Toda framework
(cf. also \cite{cb,co,ea}) and we adapt this to the example
in \cite{gc} to obtain similar relations plus a number of new formulas
and some richer perspective.  Another interesting matter involves
relations between WDVV theories for $F^{SW}$ and $F^W$; this was
discussed briefly in \cite{cb,co} and, using \cite{ib}, will be
developed further in \cite{cz} (cf. also \cite{By,ba,Bz,dc,ka,
kc,ko,mr,Mw,Ms,Mz}).

\section{THE $SU(2)$ EXAMPLE - TODA FORM}
\renewcommand{\theequation}{2.\arabic{equation}}\setcounter{equation}{0}

One goes back to \cite{sd} for this (cf. also \cite{bf,ea,gc,gf,ia,kz,la,ma,
na,se} for example - many other sources are omitted here to keep the
bibliography finite).  Thus without going into the physics here we follow
\cite{co,gf,na} at first and look at the curve ($g=N-1=1$)
\be
P(\lambda)=\Lambda^2\left(w+\frac{1}{w}\right);\,\,y=\Lambda^2\left(
w-\frac{1}{w}\right);\,\,y^2=P^2-4\Lambda^4
\label{1}
\ee
with $(\bullet)\,\,w=(1/2\Lambda^2)(P+y)$ and $w^{-1}=(1/2\Lambda^2)
(P-y)$; also ($u=u_2$)
\be
P(\lambda)=\lambda^2-u=(\lambda-\lambda_1)(\lambda-\lambda_2);\,\,u=
\lambda_1\lambda_2;
\label{2}
\ee
$$log\left(1-\frac{u}{\lambda^2}\right)=-\sum_2^{\infty}\frac{h_k}
{\lambda^k}\Rightarrow h_2=h=u$$
Further on a given curve ($u$ and $\Lambda$ fixed)
\be
dS_{SW}=\lambda\frac{dw}{w}=\frac{\lambda dy}{P}=\frac{\lambda dP}{y}=
\frac{\lambda P'd\lambda}{y}=\frac{2\lambda^2d\lambda}{y};
\label{3}
\ee
$$a=\oint_AdS_{SW}=\oint_A\frac{\lambda P'd\lambda}{\sqrt{P^2-4\Lambda^4}}=
\oint_A\frac{2\lambda^2d\lambda}{\sqrt{\lambda^4-2u\lambda^2+u^2
-4\Lambda^4}}$$
while from $\delta P+P'\delta\lambda=0$ with $\delta P=-\delta u$ one
has ($\delta w=0=\delta\Lambda$)
\be
\left.\frac{\partial dS_{SW}}{\partial u}\right|_{w=c}=\frac{\partial
\lambda}{\partial u}\frac{dw}{w}=\frac{dw}{P'w}=\frac{d\lambda}{y}=d\hat{v}
\label{4}
\ee
leading to $(\spadesuit)\,\,\partial a/\partial u=\oint_Ad\hat{v}=\sigma$
and we take $d\omega=(1/\sigma)d\hat{v}$ as the canonical holomorphic
differential.
More generally from $(\bullet)$ one has
\be
\delta P+P'\delta\lambda=NP\delta\,log(\Lambda) +y\frac{\delta w}{w}
\label{5}
\ee
\indent {\bf REMARK 2.1.}$\,\,$
Let us show how two relevant standard representations of an elliptic
curve are related (cf. \cite{kz,km}). 
Thus consider $(\clubsuit)\,\,
y^2 =(x^2-u)^2-\Lambda^4$ with $u>\Lambda$
which can be written in the form $y^2=\prod_1^2(x-e_i^{\pm})$ where
$e_1^{+}=-\sqrt{u+\Lambda^2},\,\,e_1^{-}=-\sqrt{u-\Lambda^2},$ and
$e_2^{+}=\sqrt{u+\Lambda^2}$ with $e_2^{-}=\sqrt{u-\Lambda^2}$
(note that a factor of 4 in $4\Lambda^2$ is omitted here from
(\ref{1}) - it will be inserted below when needed).
One can then produce a transformation $x\to[(ax+b)/(cx+d)]$ taking 
$e_1^{+}\to\infty,\,\,e_1^{-}\to -\Lambda^2,$ and $e_2^{-}\to
\Lambda^2$ yielding an equivalent
curve with branch points 
$\pm\Lambda^2,\,\,\tilde{u}(u,\Lambda)$, and $\infty$.
Such a curve can also be described via
$y^2=4(x-\Lambda^2)
(x+\Lambda^2)(x-\tilde{u})$ which corresponds to a curve
$y^2=4x^3-g_2(u)x-g_3(u)$ related to the Weierstrass ${\cal P}$ function
\be
{\cal P}(z,\tau)=\frac{1}{z^2}+\sum_{w\in L/\{0\}}\left(
\frac{1}{(z-w)^2}-\frac{1}{w^2}\right)
\label{7}
\ee
with $x={\cal P}$ and $y={\cal P}'$.  Here $L$ is a lattice with basic
parameters $w_1=\pi$ and $w_2=\pi\tau$ where $w_2/w_1=\tau$.
It will be convenient now to express a version of
the above calculation as follows.
Write $Q(x)=(x-a)(x-b)(x-c)(x-d)$ and set $x=a+(1/z)$ to get
\be
Q(x)=\frac{(a-b)(a-c)(a-d)}{z^4}\tilde{Q}(z);\,\,
\tilde{Q}(z)=\left(z+\frac{1}{a-b}\right)\left(z+\frac{1}{a-c}\right)
\left(z+\frac{1}{a-d}\right)
\label{8}
\ee
It follows that $(dx/\sqrt{Q(x)})=-(dz/\sqrt{\tilde{Q}(z)})$ up to a
constant multiplier.  If now we want $e_1^{+}=a\to\infty$
then write $b=e_1^{-},\,\,c=e_2^{+},$ and $d=e_2^{-}$ so in particular
$a-c=-2\sqrt{u+\Lambda^2}$.
Thus $a-b=\sqrt{u-\Lambda^2}-\sqrt{u+\Lambda^2}$ with 
$a-d=-\sqrt{u+\Lambda^2}-\sqrt{u-\Lambda^2}$.
We now map $1/(a-b)\to\Lambda^2$ and $1/(a-d)\to -\Lambda^2$ via
\be
z\to\zeta =-\frac{2\Lambda^4z}{(u-\Lambda^2)^{1/2}}-\Lambda^2\left(\frac
{u+\Lambda^2}{u-\Lambda^2}\right)^{1/2}
\label{9}
\ee
This leads to 
\be
y^2=Q(x)=\frac{(a-b)(a-c)(a-d)}{z^4}(\zeta+\Lambda^2)(\zeta+\Lambda^2)
(\zeta-\tilde{u})
\label{10}
\ee
for some $\tilde{u}=\tilde{u}(u,\Lambda)$ and then scaling via
$z^4y^2=\eta^2(a-b)(a-c)(a-d)$ we obtain
$(\bullet\bullet)\,\,\eta^2=(\zeta+\Lambda^2)(\zeta-\Lambda^2)(\zeta-
\tilde{u})$.
Thus we can work with $(\clubsuit)\,\,y^2=(x^2-u)^2-\Lambda^4$ 
(Toda form) or
with $(\bullet\bullet)\,\,\eta^2=(\zeta+\Lambda^2)(\zeta-\Lambda^2)(\zeta
-\tilde{u})$ (KdV form). 
Mathematically they would appear to be equivalent
formulations but physically one is interchanging electric and magnetic
sectors (cf. \cite{km,sd}).  For the curves $(\clubsuit)$ one
has cycles $A$ surrounding $e_1^{+}$ and $e_1^{-}$ and $B$ surrounding
$e_1^{-}$ and $e_2^{-}$ with e.g. $\oint_A\sim 2\int_{e_1^{+}}^{e_1^{-}}$.
Thus it is probably easier to work with $(\bullet\bullet)$ for the 
pure $SU(2)$ theory although for general $N_f$ and $N_c$ one has a
canonical Toda formulation involving curves such as $(\clubsuit)$.
\\[3mm]\indent
Let us now give a description of the curve $(\clubsuit)$ in
different parametrizations following \cite{co,gf} but we now insert
$\Lambda^2$ in the formulas; the calculations are worth seeing for
comparison with the modified treatment below for the curve in
KdV form. 
We write $z\sim
\xi^{-1}$ as a local coordinate near $\infty_{\pm}$ where $\infty_{\pm}$
refers to $\lambda\to\infty$ (which is not a branch point) and
recall again $(\clubsuit)\,\,y^2=(\lambda^2-u)^2-\Lambda^4=\prod_1^2
(\lambda-e_i^{\pm})$, so $\infty_{\pm}\sim (y,\lambda)=(\pm,\infty)$.
From $(\bullet)\,\,w=(1/2\Lambda^2)(P+y)$ and $w^{-1}=
(1/2\Lambda^2)(P-y)$
one sees that $w^{\pm 1}=(1/2\Lambda^2)(P\pm y)\sim 
(1/\Lambda^2)P(\lambda)(1+
O(\lambda^{-4})$ near $\infty_{\pm}$.  To see this note 
$P(\lambda)=O(\lambda^2)$ and $w^{\pm 1}=
(1/2\Lambda^2)P(\lambda)(1\pm (y/P))$.  Then
from $y^2=P^2-4\Lambda^4$ one has $(y^2/P^2)=1-(4\Lambda^4/P^2)=1-
O(\lambda^{-4})$ which implies that $y/P=1+O(\lambda^{-4})$.
Consequently one has $w^{\pm 1}\sim (1/\Lambda^2)
P(\lambda)$ near $\infty_{\pm}$ and
we note $w(\lambda=\infty_{+})=\infty$ with $w(\lambda=\infty_{-})=0$.
The curve is clearly hyperelliptic in the $(y,\lambda)$ parametrization
but not in the $(w,\lambda)$ parametrization $(w+w^{-1})\Lambda^2
=P(\lambda)=\lambda^2-u$ where $\Lambda\xi\equiv w^{-1/2}\sim
\Lambda\lambda^{-1}$
near $\infty_{+}$ and $\Lambda\xi\equiv w^{1/2}\sim
\Lambda\lambda^{-1}$ at
$\infty_{-}$ (i.e. $\Lambda\xi\sim w^{\mp 1/2}$ at $\infty_{\pm}$).
Now one defines differentials $d\Omega_n\sim 
\pm w^{\pm n/2}(dw/w)=(2/n)dw^{\pm n/2}$ near $\infty_{\pm}$
corresponding to $d\Omega_n\sim 
(2/n)d(\Lambda\xi)^{-n}$ near $\infty_{\pm}$
so $d\Omega_n\sim -2\Lambda^{-n}\xi^{-n-1}d\xi$ with normalization
$\oint_Ad\Omega_n=0$.
In taking residues at $\xi=0$ one has two points $\infty_{\pm}$ to consider
where $\Lambda\xi\sim w^{\mp 1/2}$ so $\xi^{-1}\sim P^{1/2}$ in the
$w$ parametrization and $d\Omega_n$ has two poles.  One can also look at
the hyperelliptic parametrization $(y,\lambda)$ with two sheets and talk
about $d\Omega_n^{\pm}$ based on $\xi\sim\lambda^{-1}$ with a single
pole at $(\pm,\infty)$.  We also consider the differentials
(cf. \cite{cz,ta} for clarification)
\be
d\hat{\Omega}_n=P^{n/2}_{+}(\lambda)\frac{dw}{w}
\label{11}
\ee
having poles at $\xi\sim w^{\mp 1/2}$ corresponding to $\infty_{\pm}$ 
which are to balance those of $d\Omega_n$ in a formula
($d\Omega_n=d\hat{\Omega}_n-(\oint_Ad\hat{\Omega}_n)d\omega$)
\be
dS=\sum T_nd\hat{\Omega}_n=\alpha d\omega+\sum T_nd\Omega_n;\,\,\frac
{\partial dS}{\partial \alpha}=d\omega;
\,\,\frac{\partial dS}{\partial T_n}
=d\Omega_n
\label{12}
\ee
where $\alpha$ and $T_n$ can be taken to be independent (cf. \cite{cc,ia}).
We note that these $d\Omega_n$ do not 
coincide with those of \cite{cc,na} (nor
do the $T_n$) but essentially we only deal with $T_1$ and $d\Omega_1\sim
d\Omega_1^{+}+d\Omega_1^{-}$ here and these will be the same.  In fact
we work entirely in the KdV framework starting in Section 3 and 
primarily with a standard $T_1$ and $d\Omega_1$ as in \cite{cc,na};
a $T_3$ and $d\Omega_3$ are indicated in 
passing and these are made explicit.
Now from $P_{+}^{1/2}(\lambda)=\lambda$ we see that 
$d\hat{\Omega}_1=dS_{SW}$
and near $\infty_{\pm}$ one has $d\hat{\Omega}_1\sim\mp (2d\xi/\xi)$ via
$\lambda\sim \xi^{-1}$ and $d\xi=\mp (1/2)w^{\mp (1/2)-1}dw$.  Thus
in particular this is compatible with 
$d\Omega_1\sim\pm w^{\pm 1/2}(dw/w)\sim
\pm(\Lambda\xi)^{-1}(dw/w)$, or more generally 
$d\Omega_n\sim\pm (2/n)d(\Lambda\xi)^{-n}\sim
\mp 2\Lambda^{-n}(d\xi/\xi^{n+1})$.  
Now from (\ref{12}) we obtain the Whitham dynamics
of $u$ from
\be
\frac{\partial dS}{\partial T_n}=d\hat{\Omega}_n+\sum T_m\frac{\partial
d\hat{\Omega}_m}{\partial u}\frac{\partial u}{\partial T_n}=
d\Omega_n\Rightarrow
\label{13}
\ee
$$\Rightarrow -\oint_Ad\hat{\Omega}_n=\sum 
T_m\frac{\partial u}{\partial T_n}
\oint_A\frac{\partial d\hat{\Omega}_m}{\partial u}$$
since $\oint_Ad\Omega_n=0$.  But a basic assumption is 
that $\partial dS/
\partial u=dV$ where $dV$ is a holomorphic differential
(which can be achieved via a stipulation
$\partial d\hat{\Omega}_n/\partial u=\beta_nd\omega$), and of necessity
$d\hat{\Omega}_n=d\Omega_n+c_nd\omega$ (where 
$\oint_Ad\hat{\Omega}_n
=c_n$).  Thus
$\oint_A(\partial d\hat{\Omega}_m/\partial u)=\oint_A
\beta_md\omega=\beta_m$
and (\ref{13}) implies $(\clubsuit\clubsuit)\,\,\partial u/\partial T_n
=-c_n(\sum T_m\beta_m)^{-1}$.
\\[3mm]\indent
Now the procedure of \cite{gf} (reproduced in
\cite{co}) gives an abstract formula ($u\sim u_2\sim h$)
\be
\left.\frac{\partial u}{\partial\,log(\Lambda)}\right|_{a=c}=
2u-a\frac{\partial u}{\partial a}
\label{16}
\ee
(cf. below) as well as an expression for $\partial u/\partial\,log
(\Lambda)$ in terms of theta functions (which we omit here).
To see how this goes
we recall first (cf. \cite{co,gf}) an abstract derivation of connections
between $log(\Lambda)$ and $log(T_1)$ derivatives.  Thus first
note from (\ref{12}) that
\be
\oint_{A_i}dS=\sum T_n\oint_{A_i}d\hat{\Omega}_n=
\alpha_i\Rightarrow \alpha_i=
T_1a_i+O(T_2,T_3,\cdots)
\label{26}
\ee
Hence for our example $\alpha=T_1a+O(T_2,T_3,\cdots)$.  Next write from
(\ref{1})
\be
\delta P+P'\delta\lambda=2P\delta\,log(\Lambda)+y\frac{\delta w}{w}
\label{27}
\ee
so for $\delta w=0$ one has $(\spadesuit\spadesuit)\,\,P'\delta\lambda-
\delta u=2\lambda\delta\lambda-\delta u=2P\delta\,log(\Lambda)$.  Also
note from $P=\lambda^2-u$ one has $(\bullet\bullet\bullet)\,\,
\lambda dP=\lambda[2\lambda d\lambda]=2(P+u)d\lambda$.  Hence from
$(\spadesuit\spadesuit)$ for $\delta w=0$
\be
\delta a=\oint_A\delta\lambda\frac{dw}{w}=\delta u\oint_A\frac{dw}
{2\lambda w}+\delta\,log(\Lambda)\oint_A\frac{2P}{P'}\frac{dw}{w}
\label{28}
\ee
Recall now $dv=(1/P')(dw/w)=(d\lambda/y)$ and $\partial a/\partial u=
\oint_Adv=\sigma$ so for $\delta a=0$ one obtains
$$
\left(\left.
\frac{\partial u}{\partial\,log(\Lambda)}\right|_{a=c}\right)
\oint_Adv=-\oint_A\frac{2P}{P'}\frac{dw}{w}
=-\oint_A\frac{2Pd\lambda}{y}=$$
\be
=-\oint_A\frac{\lambda dP-2ud\lambda}{y}=-a+2u\oint_Adv
\label{29}
\ee
Thus, multiplying by $\partial u/\partial a=1/\sigma$
we get (\ref{16}). 
Next from (\ref{26}) one has
\be
\delta\alpha=a\delta T_1+T_1\delta a+O(T_2,T_3,\cdots)\Rightarrow
\label{30}
\ee
$$\Rightarrow \delta\alpha=\alpha\delta\,log(T_1)+T_1\oint_A\delta\lambda
\frac{dw}{w}+O(T_2,T_3,\cdots)$$
Hence for constant $\Lambda$ and $T_n=0$ for $n\geq 2$ (with $\alpha$ and
$T_n$ independent) $\delta\alpha=0$ implies
$$
\frac{\alpha}{T_1}=-\oint_A\frac{\partial\lambda}{\partial\,log(T_1)}
\frac{dw}{w}=-\oint_A\frac{\partial\lambda}{\partial u}
\frac{\partial u}{\partial\,log(T_1)}\frac{dw}{w}=$$
\be
=-\frac{\partial u}{\partial\,log(T_1)}\oint_Adv=-\sigma\left.\frac
{\partial u}{\partial\,log(T_1)}\right|_{\alpha=c}
\label{31}
\ee
(cf. (\ref{4}) and $(\spadesuit)\,\,\partial a/\partial u=\sigma$).
Thus we have $-\sigma(\partial u/\partial\,log(T_1))=a$ from (\ref{31})
and $\sigma(\partial u/\partial\,log(\Lambda))=-a+2u(\partial a/
\partial u)$ from (\ref{29}), which implies
\be
\left.
\frac{\partial u}{\partial\,log(T_1)}\right|_{\alpha=c}
=-\frac{a}{\sigma}=-a\frac{\partial u}
{\partial a}=\left.
\frac{\partial u}{\partial\,log(\Lambda)}\right|_{a=c}
-\frac{2u}{\sigma}
\frac{\partial a}{\partial u}=\left.
\frac{\partial u}{\partial\,log(\Lambda)}\right|_{a=c}-2u
\label{32}
\ee
\indent {\bf REMARK 2.2.}$\,\,$
Note that $\partial d\hat{\Omega}_1/\partial u=\beta_1d\omega=dv$ as in
(\ref{4}) and $d\hat{\Omega}_1=d\Omega_1+c_1d\omega$ (cf. remarks
after (\ref{13})) imply $(\clubsuit\clubsuit\clubsuit)\,\,\sigma=
\oint_Adv=\beta_1$ and $c_1=\oint_Ad\hat{\Omega}_1=a$.  Hence the Whitham
dynamics in $(\clubsuit\clubsuit)$ for $u$ become for $T_n=0\,\,(n\geq 2)$
\be
\frac{\partial u}{\partial T_1}=-c_1(T_1\beta_1)^{-1}=-\frac{a}{T_1\sigma}
=-\frac{a}{T_1}\frac{\partial u}{\partial a}
\label{33}
\ee
(recall $\sigma=\partial a/\partial u$) and this is precisely the 
homogeneity condition for moduli:  $T_1\partial_1u+a\partial_au=0$ (cf.
\cite{cc,ia}).
\\[3mm]\indent {\bf REMARK 2.3.}$\,\,$  The equations 
\be
\frac{\partial F^W}{\partial T_1}=\frac{2T_1h_2}{i\pi};\,\,\Lambda
\partial_{\Lambda}F^{SW}=\frac{2T_1^2u}{i\pi}
\label{34}
\ee
for $T_n=0\,\,(n\geq 2)$, from \cite{co,gf} (with $\Lambda\sim T_1$ in
the SW theory) should correspond here to an equation
$\partial F^{SW}/\partial\,log(\Lambda)=
\partial F^W/\partial\,log(T_1)$.  This involves 
a heavy abuse of notation
since $F^W\to F^{SW}$ when $T_n\to\delta_{n,1}$ and thus $\Lambda$ seems
to acquire a fixed value $1$ if $\Lambda\sim T_1$.  
Note in \cite{cb,co,ea}
for example that $T_1$ acquires a fixed value $1/i\pi$ for pure $SU(2)$
SW theory in a slightly different normalization 
($N_f=0,\,\,N_c=2,\,\,m_i=0$).  
The comment in \cite{gf} in this regard is that the SW theory 
itself allows for 
$T_1$ derivatives since after rescaling $h_k\to 
T_1^kh_k$ and ${\cal H}_k\to
T_1^k{\cal H}_k$ one can identify $T_1$ with $\Lambda$, and $\Lambda$ is
explicitly present in the SW theory (which is rather confusing).  
We will clarify all this below
for the curve in KdV form and similar considerations apply for the 
Toda form.

\section{EXPLICIT FORMULAS FOR THE KdV CURVE}
\renewcommand{\theequation}{3.\arabic{equation}}\setcounter{equation}{0}

We consider the curve now in the form (cf. Remark 2.1) 
$y^2=(\lambda-\Lambda^2)(\lambda+\Lambda^2)(\lambda-v)$ as in
\cite{bc,ba,Bz,gc,ma,sd}.
First we want an explicit connection between formulas based on the curve
in two equivalent forms ${\bf (A)}\,\,y^2=(\lambda-\Lambda^2)(\lambda
+\Lambda^2)(\lambda-v)$ and ${\bf (B)}\,\,\eta^2=(\mu-1)(\mu+1)
(\mu-\tilde{v})$.  Thus in {\bf (A)} set $\lambda=\Lambda^2\mu$ to get
$y^2=\Lambda^6(\mu-1)(\mu+1)(\mu-\tilde{v})$ for $\tilde{v}=v/\Lambda^2$.
Then set $y^2=\Lambda^6\eta^2$ for the form {\bf (B)}.  Compare then e.g.
the formulas
\be
a(v,\Lambda)=\frac{\sqrt{2}}{\pi}\int_{-\Lambda^2}^{\Lambda^2}
\frac{d\lambda\sqrt{\lambda-v}}{\sqrt{\lambda^2-\Lambda^4}};\,\,\tilde
{a}(\tilde{v})=\frac{\sqrt{2}}{\pi}\int_{-1}^1
\frac{d\mu\sqrt{\mu-\tilde{v}}}
{\sqrt{\mu^2-1}}
\label{41}
\ee
from \cite{ba} and \cite{ma} respectively where e.g.
$dS_{SW}=(1/\pi\sqrt{2})[(\lambda-v)d\lambda/y]$ with $\partial dS_{SW}/
\partial v=-(1/2\pi\sqrt{2})(d\lambda/y)$. 
Setting $\lambda=\Lambda^2\mu$
again one obtains $a(\Lambda,v)\sim\Lambda\tilde{a}
(\tilde{v}=(v/\Lambda^2))$
(which corresponds to $\Lambda b(\tilde{v})$ in \cite{ba}), and similarly
$a^D(v,\Lambda)\sim\Lambda\tilde{a}^D(\tilde{v}=(v/\Lambda^2))$
where $a^D(v,\Lambda)=(\sqrt{2}/\pi)\int_{\Lambda^2}^v[\sqrt{\lambda
-v}/\sqrt{\lambda^2-\Lambda^4}]d\lambda$.
We recall also the Picard-Fuchs type equations 
\be
\left[(1-\tilde{v}^2)\partial^2_{\tilde{v}}-\frac{1}{4}\right]
\left(
\begin{array}{c}
\tilde{a}\\
\tilde{a}^D
\end{array}\right)=0\equiv
\left[\partial^2_v+\frac{1}{4(v^2-\Lambda^4)}\right]\left(
\begin{array}{c}
a\\
a^D
\end{array}\right)=0
\label{43}
\ee
In \cite{ba} one concludes that $aF_a^{SW}+
\Lambda F_{\Lambda}^{SW}=2F^{SW}$
since $\tau=F_{aa}^{SW}$ is dimensionless and thence from $2F^{SW}-
aF_a^{SW}=-8\pi iv/4\pi^2=-(2iv/\pi)$ that $\Lambda F_{\Lambda}^{SW}=
-(2iv/\pi)$ (cf. below for more detail). 
Further, beta functions are defined in \cite{ba} via
\be
\beta(\tau)=\left.\Lambda\partial_{\Lambda}\tau\right|_{v=c};\,\,
\beta^a(\tau)=\left.\Lambda\partial_{\Lambda}\tau\right|_{a=c}
\label{44}
\ee
Some calculation gives then for $G(\tau)=v/\Lambda^2$ the formula
$\Lambda\partial_{\Lambda}G=-2v/\Lambda^2$ so $\Lambda\partial_{\Lambda}
G=\Lambda\partial_{\tau}G\partial_{\Lambda}\tau=
\beta(\tau)\partial_{\tau}G=
-2v/\Lambda^2=-2G$ which implies $\beta(\tau)=-2G/G_{\tau}$ and
$G(0)=1$ can be shown.  This leads to
\be
v=\Lambda^2exp\left(-2\int_0^{\tau}\beta^{-1}(x)dx\right)\equiv
v(\tau)=\left(\frac{\Lambda}{\Lambda_0}\right)^2v(\tau_0)
e^{-2\int_{\tau_0}^{\tau}\beta^{-1}(x)dx}
\label{45}
\ee
Then from $2F^{SW}-aF^{SW}_a=-(2iv/\pi)=\Lambda F_{\Lambda}^{SW}$ one has
\be
\partial_{\Lambda}F^{SW}=-\frac{2iv\Lambda}{\pi}
e^{-2\int_0^{\tau}\beta^{-1}
dx}\equiv \partial_{\Lambda}F^{SW}(a,\Lambda)=\frac{\Lambda}{\Lambda_0}
\partial_{\Lambda_0}F^{SW}(a_0,\Lambda_0)e^{-2\int_{\tau_0}^{\tau}
\beta^{-1}dx}
\label{46}
\ee
Another expression for $\beta$ arises from (cf. \cite{ba} and recall
$a(\Lambda,v)\sim\Lambda b(\tilde{v})$ with $\tilde{v}=v/\Lambda^2$)
\be
d\tau=(\partial_{\Lambda}\tau)_ad\Lambda+(\partial_a\tau)_{\Lambda}da
=(\partial_{\Lambda}\tau)_vd\Lambda+(\partial_v\tau)_{\Lambda}dv;
\label{47}
\ee
$$da=(\partial_{\Lambda}a)_vd\Lambda+(\partial_va)_{\Lambda}dv=
bd\Lambda+\Lambda b'd\tilde{v}=(b-2\tilde{v}b')d\Lambda+
\Lambda^{-1}b'dv$$
This leads to
\be
(\partial_{\Lambda}\tau)_v=(\partial_{\Lambda}\tau)_a+(b-2\tilde{v}b')
(\partial_a\tau)_{\Lambda}
\label{48}
\ee
But moduli are homogeneous functions of degree zero, i.e.
\be
a(\partial_a\tau)_{\Lambda}+\Lambda(\partial_{\Lambda}\tau)_a=0
\label{49}
\ee
This can be seen in various ways (cf. \cite{ba,ia}) and is shown explicitly
below.  
Now using (\ref{49}) in (\ref{48}) one obtains (recall $a=\Lambda b$)
\be
\beta(\tau)=\Lambda(\partial_{\Lambda}\tau)_v=-a(\partial_a\tau)_{\Lambda}
+(b-2\tilde{v}b')(\partial_a\tau)_{\Lambda}=-2\Lambda\tilde{v}b'(\partial_a
\tau)_{\Lambda}=\frac{2\tilde{v}b'}{b}\beta^a(\tau)
\label{51}
\ee
since $\beta^a=\Lambda(\partial_{\Lambda}\tau)_a=
-a(\partial_a\tau)_{\Lambda}
=-\Lambda b(\partial_a\tau)_{\Lambda}$ from (\ref{44}).  Note also 
$\Lambda(\partial_{\Lambda}\tau)_a=-a(\partial_a\tau)_{\Lambda}=
-\Lambda b(\partial_b\tau/\Lambda)_{\Lambda}=-b(\partial_b\tau)_{\Lambda}$.
\\[3mm]\indent
Let us go now to \cite{gc} and rewrite some of the formulas with 
$\Lambda^2$ inserted.  First from \cite{bf} the notation will involve
$u\to\tilde{v}$ and $x\sim z\to\mu$ with
$$
a=\Lambda\tilde{a}=\Lambda\sqrt{2}(\tilde{v}+1)^{1/2}F\left(-\frac{1}{2},
\frac{1}{2},1,\frac{2}{\tilde{v}+1}\right)=\sqrt{2}(v+\Lambda^2)^{1/2}
F\left(-\frac{1}{2},\frac{1}{2},1,\frac{2\Lambda^2}{v+\Lambda^2}\right);$$
\be
a^D=\Lambda\tilde{a}^D=\frac{i}{2\Lambda}(v-\Lambda^2)^{1/2}F\left(
\frac{1}{2},\frac{1}{2},2,\frac{\Lambda^2-v}{2\Lambda^2}\right)
\label{55}
\ee
Now as in \cite{gc}, first in the $(z,u)$ notation, one 
can look at the elliptic
curve as a finite zone KdV curve (e.g. think of 
an elliptic 1-gap solution to
KdV based on the Weierstrass ${\cal P}$ function) so that there can
be a background KdV theory associated to the curve (cf. \cite{bh,cc,cs,fb,
nb} for more detail).
Here one recalls that for KdV there are differentials 
$
d\Omega_{2j+1}(z)=(P_{j+g}(z)/y(z))dz;\,\,y^2\sim (z^2-1)(z-u)$.
In particular one writes
\be
dp\equiv d\Omega_1=\frac{z-\alpha(u)}{y(z)}dz;\,\,dE\equiv d\Omega_3(z)
=\frac{z^2-\frac{1}{2}uz-\beta(u)}{y(z)}dz
\label{FM}
\ee
The normalization conditions $\oint_A d\Omega_i=0$ yield
$\alpha(u)$ and $\beta(u)$ immediately.  Associated with this situation
we have the classical Whitham theory (cf.
\cite{bb,cc,cg,dc,fb,ka,kc,ko,na}) giving ($t_n\to T_n=
\epsilon t_n,\,\,\epsilon\to 0$)
$
(\partial d\Omega_i(z)/\partial T_j)=(\partial d\Omega_j(z)/
\partial T_i);\,\,d\Omega_i(z)=(\partial dS(z)/\partial T_i)$
where $dS$ is some action term which classically was thought of
in the form $dS=\sum T_id\Omega_i$.  In fact it will continue to have
such a form in any context for generalized times $T_A$ (including the
$a_i$) and generalized differentials $d\Omega_A$ (cf. \cite{cb,kc}).  
It is shown
in \cite{gc} that 
\be
dS(z)=\left(T_1+T_3(z+\frac{1}{2}u)+\cdots\right)\times
\frac{z-u}{y(z)}dz=g(z|T_i,u)\lambda_{SW}(z)
\label{FPP}
\ee
where $\lambda_{SW}$ is the SW differential $(z-u)dz/y(z)$
(actually $\lambda_{SW}=\pi\sqrt{2}dS_{SW}$ - cf. (\ref{41})).
The demonstration is sort of ad hoc and goes as follows.  
Setting $T_{2k+1}=0$ for $k>1$ and computing from
(\ref{FPP}) one gets
\be
\frac{\partial dS(z)}{\partial T_1}=\left(z-u-
(\frac{1}{2}T_1+\frac{3}{4}uT_3)
\frac{\partial u}{\partial T_1}\right)\frac{dz}{y(z)};
\label{FQQ}
\ee
$$\frac{\partial dS(z)}{\partial T_3}=\left(z^2-
\frac{1}{2}uz-\frac{1}{2}u^2
-(\frac{1}{2}T_1+\frac{3}{4}uT_3)\frac{\partial u}
{\partial T_3}\right)\frac{dz}{y(z)}$$
so requiring $\partial_1dS=d\Omega_1$ and $\partial_3dS=d\Omega_3$ gives
via (\ref{FM})
\be
(\frac{1}{2}T_1+\frac{3}{4}uT_3)\frac{\partial u}
{\partial T_1}=\alpha(u)-u;\,\,(\frac
{1}{2}T_1+\frac{3}{4}uT_3)\frac{\partial u}
{\partial T_3}=\beta(u)-\frac{1}{2}u^2
\label{FR}
\ee
Hence the construction gives a solution to the general Whitham equation
of the form ${\bf (WW)}\,\,
\partial u/\partial T_3=v_{31}(u)(\partial u/\partial T_1)$ with
$
v_{31}=[\beta(u)-(1/2)u^2]/[\alpha(u)-u]=
[d\Omega_3(z)/[d\Omega_1(z)]$ (evaluated at $z=u$),
which is what it should be from the general Whitham theory (cf. 
\cite{cc,fb}).
It follows that (modulo a constant)
\be
a=\left.\frac{1}{T_1}\oint_AdS(z)\right|_{T_3=T_5=\cdots=0};\,\,
a_D=\left.\frac{1}{T_1}\oint_BdS(z)\right|_{T_3=T_5=\cdots=0}
\label{FT}
\ee
Note here $\left.(1/T_1)dS\right|_{T_3=0}=[(z-u)/y(z)]dz=
\lambda_{SW}$ is fine but one does not seem to have the form 
$d{\cal S}=ad\omega+\sum T_nd\Omega_n$ as in \cite{ia}
(cf. also \cite{cc,cg}) where
$cd\omega=dz/y(z)=d\hat{v}$ is a holomorphic differential with
$\oint_Ad\omega=1$ and $\oint_Bd\omega=\tau$. 
To put this in a more canonical form one obtains $\alpha$ via
\be
\oint_Ad\Omega_1=0=\oint_A\frac{z-\alpha}{y}dz=
\oint_A\frac{zdz}{y}-\alpha c
\Rightarrow\alpha=\frac{1}{c}\oint_A\frac{zdz}{y}
\label{56}
\ee
Similarly $(1/c)\oint_A[(z^2-(1/2)uz)/y]dz=\beta$ and for $T_3=0$
\be
dS=T_1\lambda=T_1\left[\frac{z-\alpha}{y}+\frac{\alpha-u}{y}\right]dz
=T_1d\Omega_1+T_1(\alpha(u)-u)cd\omega
\label{57}
\ee
Thus $T_1(\alpha(u)-u)c\sim T_1\hat{a}\,\,(\sim\alpha$ in (\ref{12}))
where $4\hat{a}=\oint_A\lambda$ here, and we refer to (\ref{63}) for a
slightly ``refined" version.
\\[3mm]\indent
Let us now put in the constant $\pi\sqrt{2}$ and $\Lambda$ so 
$dS_{SW}=(1/\pi\sqrt{2})\lambda_{SW}$ with $a=\oint_AdS_{SW}=
(1/\pi\sqrt{2})\oint_A\lambda_{SW}$ 
as in (\ref{41}) and we set $u=v/\Lambda^2$ with
$z=\lambda/\Lambda^2$.  Then $\Lambda^6y^2=(\lambda^2-\Lambda^4)(\lambda
-v)=\eta^2$ (here $\eta\leftrightarrow y$ in {\bf (B)}) with 
$dz=\Lambda^{-2}d\lambda$ and
\be
d\Omega_1=\frac{(\lambda-\tilde{\alpha})d\lambda}{\Lambda\eta}=\frac
{1}{\Lambda}d\tilde{\Omega}_1;\,\,d\Omega_3=
\frac{(\lambda^2-(1/2)v\lambda
-\tilde{\beta})d\lambda}{\Lambda^3\eta}=\frac{1}{\Lambda^3}d\tilde
{\Omega}_3
\label{58}
\ee
where $\tilde{\alpha}=\Lambda^2\alpha(v/\Lambda^2)$ and $\tilde{\beta}=
\Lambda^4\beta(v/\Lambda^2)$.  Further ${\bf (SW)}\,\,
dS_{SW} =(1/\pi\sqrt{2}(\lambda-v)d\lambda/\eta)=
(\Lambda/\pi\sqrt{2})\lambda_{SW}$
since $\lambda_{SW}=[(z-u)/y]dz=(\lambda-v)d\lambda/\Lambda\eta$.
Now we can drop the $\Lambda$ and $\Lambda^3$ factors in $d\Omega_1$ and
$d\Omega_3$ (they do not affect normalization and only contribute a
multiplier to the asymptotics) and simply take $dS_{SW}$ as in 
{\bf (SW)} with $\oint_AdS_{SW}=a$ to coincide with (\ref{41}).
Then use $\oint_Ad\tilde{\Omega}_i=0$ to get
\be
0=\oint_A\frac{\lambda-\tilde{\alpha}}{\eta}d\lambda=\oint_A\frac{\lambda
d\lambda}{\eta}-\tilde{\alpha}\oint_Ad\tilde{v}\Rightarrow\tilde{\alpha}=
\frac{1}{\tilde{c}}\oint_A\frac{\lambda d\lambda}{\eta}
\label{60}
\ee
(i.e. $d\tilde{v}=\tilde{c}d\tilde{\omega}$ where $d\tilde{\omega}$ is the
canonical holomorphic differential - note $d\hat{v}=(dz/y)=(d\lambda/
\Lambda\eta)=(1/\Lambda)d\tilde{v}$). 
Similarly ${\bf (E)}\,\,
(1/\tilde{c})\oint_A[(\lambda^2-(1/2)\lambda v)/\eta]d\lambda=
\tilde{\beta}$.  Now one can write
\be
T_1dS_{SW}=\frac{T_1}{\pi\sqrt{2}}\left[\frac{\lambda-\tilde{\alpha}}
{\eta}+\frac{\tilde{\alpha}-v}{\eta}\right]d\lambda=
\label{61}
\ee
$$=\frac{T_1}{\pi\sqrt{2}}[\tilde{c}(\tilde{\alpha}-v)d\tilde{\omega}
+d\tilde{\Omega}_1]$$
with ${\bf (EE)}\,\,a=(1/\pi\sqrt{2})\tilde{c}(\tilde{\alpha}
-v)$.  Similarly
$
T_3[z+(1/2)u]dS_{SW}$ becomes
\be
T_3
\left(\frac{\lambda+(1/2)v}
{\Lambda^2}\right)\left(\frac{(\lambda-v)d\lambda}
{\pi\sqrt{2}\eta}\right)=
\frac{T_3}{\Lambda^2\pi\sqrt{2}}\left[d\tilde{\Omega}_3+
\left(\tilde{\beta}-\frac{1}{2}v^2\right)\frac{d\lambda}{\eta}\right]
\label{62}
\ee
leading to
\\[3mm]\indent {\bf PROPOSITION 3.1.}$\,\,$  
The differential $dS$ of (\ref{FPP})
has a ``canonical" form
\be
d\tilde{S}=[T_1+T_3(z+(1/2)u)]dS_{SW}=T_1\left[ad\tilde{\omega}+\frac
{1}{\pi\sqrt{2}}d\tilde{\Omega}_1\right]+
\label{63}
\ee
$$+ \frac{T_3}{\Lambda^2\pi\sqrt{2}}\left[
d\tilde{\Omega}_3-\left(\tilde{\beta}-\frac{1}{2}v^2\tilde{c}
d\tilde{\omega}\right)
\right]$$
\indent
Apparently the only reason for the $T_3$ term in \cite{gc} was to 
exhibit the Whitham equation {\bf (WW)} but this has
basically nothing to do with SW theory as such.
In order to study relations
between $T_1$ and $\Lambda$ derivatives we set now
$T_3=0$ and look at Whitham theory based
on $T_1$ and $a$, so
\be
d\tilde{S}=T_1ad\tilde{\omega}+\frac{T_1}{\pi\sqrt{2}}d\tilde{\Omega}_1
=\gamma d\tilde{\omega}+T_1d\check{\Omega}_1
\label{64}
\ee
where $\gamma=aT_1$ is used instead of $\alpha$ as in (\ref{12})
since $\alpha,\,\tilde{\alpha}$ appear here already with other meanings.
Here $\Lambda$ and $v$ will satisfy ($\oint_Ad\tilde{\omega}
=1$)
\be
\tilde{c}(v,\Lambda)d\tilde{\omega}=d\tilde{v}=\frac{d\lambda}{\eta};\,\,
a=\frac{1}{\pi\sqrt{2}}\tilde{c}
(\tilde{\alpha}-v);
\label{65}
\ee
$$\tilde{\alpha}=\frac{1}{\tilde{c}}\oint_A\frac{\lambda d\lambda}{\eta}=
\tilde{\alpha}(v,\Lambda);\,\,d\tilde{\Omega}_1=\frac{\lambda-\tilde
{\alpha}}{\eta}d\lambda;\,\,d\check{\Omega}_1=
\frac{1}{\pi\sqrt{2}}d\tilde{\Omega}_1$$
The object now is to understand the relations between $T_1$ and $\Lambda$
derivatives related to
$F^{SW}$ and $F^W$.
\\[3mm]\indent {\bf REMARK 3.2.}$\,\,$   Note an argument analogous
to (\ref{5}), (\ref{27}), (\ref{30}), (\ref{32}) can be envisioned 
based on $\eta^2=(\lambda^2-\Lambda^4)(\lambda
-v)$ via
\be
2\eta \delta\eta =[2\lambda(\lambda-v)+(\lambda^2-\Lambda^4)]\delta\lambda
-3\Lambda^3\delta\Lambda (\lambda -v)-(\lambda^2-\Lambda^4)\delta v=
\label{66}
\ee
$$=(\lambda^2-2v\lambda-\Lambda^4)\delta\lambda-3\Lambda^3(\lambda-v)
\delta\Lambda-(\lambda^2-\Lambda^4)\delta v$$
but arguments using this in a manner similar to Section 2 seem too contrived
so we prefer to work from explicit formulas below.
\\[3mm]\indent
We note next that (cf. (\ref{4}))
\be
a=\frac{1}{\pi\sqrt{2}}\oint_A\frac{\sqrt{\lambda-v}}
{\sqrt{\lambda^2-\Lambda^4}}d\lambda;\,\,\left.
\frac{\partial a}{\partial v}\right|_{\Lambda}=
-\frac{1}{2\pi\sqrt{2}}\oint_A
\frac{d\lambda}{\eta}=-\frac{\tilde{c}(v,\Lambda)}{2\pi\sqrt{2}}
=\tilde{\sigma}
\label{67}
\ee
Similarly (considering the $A$ cycle as fixed and encircling the cut)
\be
\left.\frac{\partial a}{\partial\Lambda}\right|_v=
\frac{3\Lambda^3}{2\pi\sqrt{2}}\oint_A
\frac{\lambda-v}{\eta (\lambda^2-\Lambda^4)}d\lambda
\label{68}
\ee
This integral cannot as such be written as 
$2\int_{-\Lambda^2}^{\Lambda^2}$ and
moreover we probably should better consider 
$2\int_{-\Lambda^2}^{\Lambda^2}$ 
for $a$ directly and include variation of end points with $\Lambda$.
(in fact (\ref{68}) seems incompatible with explicit formulas as in
(\ref{133}) so we will forget it).
Here (\ref{67}) is analogous to $(\spadesuit)\,\,\partial a/\partial u
=\sigma$.
\\[3mm]\indent 
We note now from \cite{cu} that for 
$d\hat{v}=dx/y=dx/\sqrt{x(x-1)(x-v)}$ one
can write $\pi_1=2\int_0^1d\hat{v}$ with $\pi_2=2\int_1^vd\hat{v}$
(there are too many $v$, $\hat{v}$, and $\tilde{v}$ symbols 
but no confusion should arise).  Then compute as in \cite{cu}
\be
\partial_v\left(\frac{1}{y}\right)=\frac{1}{2y(x-v)};\,\,
\partial^2_v\left(\frac{1}{y}\right)=\frac{3}{4y(x-v)^2}
\label{69}
\ee
\be
d\left(\frac{y}{(x-v)^2}\right)=-\frac{1}{2}d\tilde{v}-(4v-2)\partial_v
d\tilde{v}-2v(v-1)\partial^2_vd\tilde{v}\Rightarrow
\label{70}
\ee
$$\Rightarrow \frac{1}{4}\pi_i+(2v-1)\partial_v\pi_i+v(v-1)\partial^2_v
\pi_i=0$$
These are the Picard-Fuchs (PF) equations (cf. also (\ref{43})).
Applied to our situation we can modify the calculations in \cite{cu}
and consider $\eta=\sqrt{(\lambda-\Lambda^2)
(\lambda+\Lambda^2)(\lambda-v)}$
with $d\tilde{v}=d\lambda/\eta$.  Then
\be
\partial_vd\tilde{v}=\partial_v\left(\frac{d\lambda}{\eta}\right)=
\frac{d\lambda}{2\eta (\lambda-v)};\,\,\partial^2_vd\tilde{v}=
\frac{3d\lambda}{4\eta (\lambda-v)^2}
\label{71}
\ee
Now we observe that
$$
\Xi=d\left(\frac{\eta}{(\lambda-v)^2}\right)=\frac{d\lambda}
{2(\lambda-v)^{3/2}}\left((\lambda-\Lambda^2)^{1/2}(\lambda+\Lambda^2)^
{-1/2}+(\lambda+\Lambda^{1/2}(\lambda-\Lambda^2)^{-1/2}\right)+$$
\be
+\frac{d\lambda}{2(\lambda-v)^{3/2}}\sqrt{\lambda^2-\Lambda^4}-
\frac{2\eta d\lambda}{(\lambda-v)^3}
\label{72}
\ee
and this becomes
\be
\Xi=d\lambda\left[\frac{(\lambda+\Lambda^2)+(\lambda-\Lambda^2)}
{2\eta(\lambda-v)}-\frac{3(\lambda^2-\Lambda^4)}{2\eta (\lambda-v)^2}
\right]=
\label{722}
\ee
$$=(\lambda+\Lambda^2)\partial_vd\tilde{v}+(\lambda-\Lambda^2)
\partial_vd\tilde{v}-2(\lambda^2-\Lambda^4)\partial^2_vd\tilde{v}=
(\lambda-v+v+\Lambda^2)\partial_vd\tilde{v}+$$
$$+(\lambda-v+v-\Lambda^2)\partial_vd\tilde{v}-2(\lambda-v+v+\Lambda^2)
(\lambda-v+v-\Lambda^2)\partial^2_vd\tilde{v}=$$
$$=\frac{d\tilde{v}}{2}+(v+\Lambda^2)\partial_vd\tilde{v}+
\frac{d\tilde{v}}{2}+(v-\Lambda^2)\partial_vd\tilde{v}-2(\lambda-v+v+
\Lambda^2)\left(\frac{3}{2}\partial_vd\tilde{v}+(v-\Lambda^2)\partial^2_v
d\tilde{v}\right)=$$
$$=d\tilde{v}+2v\partial_vd\tilde{v}-\frac{3}{2}d\tilde{v}-3(v+\Lambda^2)
\partial_vd\tilde{v}-3(v-\Lambda^2)\partial_vd\tilde{v}-2(v^2-\Lambda^4)
\partial^2_vd\tilde{v}=$$
$$=-\frac{1}{2}d\tilde{v}-4v\partial_vd\tilde{v}-2(v^2-\Lambda^4)
\partial^2_vd\tilde{v}$$
\indent
{\bf PROPOSITION 3.3.}$\,\,$  Integrating in (\ref{722}) gives the PF
equations ($d\tilde{v}=d\lambda/\eta$)
\be
0=\frac{1}{4}\pi_i+2v\partial_v\pi_i+(v^2-\Lambda^4)\partial^2_v\pi_i
\label{73}
\ee
where $\pi_1=2\int_{-\Lambda^2}^{\Lambda^2}d\tilde{v}\sim\oint_A
d\tilde{v}=\tilde{c}=-2\pi\sqrt{2}\tilde{\sigma}$
and $\pi_2=2\int_{\Lambda^2}^vd\tilde{v}\sim\oint_Bd\tilde{v}$.
\\[3mm]\indent {\bf REMARK 3.4.}$\,\,$  In particular we have
\be
\frac{1}{4}\tilde{c}+2v\partial_v\tilde{c}+(v^2-\Lambda^4)\partial^2_v\tilde
{c}=0\equiv 2v\partial_v
\tilde{\sigma}+(v^2-\Lambda^4)\partial_v^2\tilde{\sigma}+
\frac{1}{4}\tilde{\sigma}=0
\label{74}
\ee
for comparison to 
a differentiated form of the Picard-Fuchs equations from \cite{gf}, namely
${\bf (PF)}\,\,
4(4\Lambda^4-h^2)(\partial\sigma/\partial h)=a$ (cf. also (\ref{43})).
In fact this is exact (modulo some parameter adjustment)
since $\partial a/\partial v
=\tilde{\sigma}$ while in ${\bf (PF)}\,\, \partial a/\partial h=\sigma$.
Consequently one can use heuristically
\be
4(\Lambda^4-v^2)\partial_v\tilde{\sigma}=a
\label{734}
\ee
as an integrated form of (\ref{74}) (cf. \cite{ba,ma} to confirm
exactness).
\\[3mm]\indent {\bf REMARK 3.5.}$\,\,$  Let us remark that from
Whitham theory with $d\tilde{S}$ as in (\ref{63}) one will have
$(\Lambda,v)$ as functions of $(\gamma,T_1)$
with $\gamma,\,\,T_1$ regarded as independent (cf. \cite{cc,ia}).
However since $a=a(v,\Lambda)$ one can think of $a$ as a function
of $(\gamma,T_1)$ or think of $v=v(a,\Lambda)$, or introduce $\tau
=\tau(v,\Lambda)$ as a modulus with perhaps $\tau=\tau(a,\Lambda)$,
etc. (cf. \cite{ba,ma} and (\ref{41}) - (\ref{51})).  Thus we
will try to specify the fixed variables in partial derivatives
when any confusion can arise,
e.g. $(\partial_{\Lambda}\tau)_v\equiv\partial_{\Lambda}\tau|_v$ etc.

\section{PREPOTENTIALS}
\renewcommand{\theequation}{4.\arabic{equation}}\setcounter{equation}{0}

Let us try to construct the prepotentials $F^{SW}$ and $F^W$ now for
the curve of Section 3.  For the Whitham theory we can use the
formulation in \cite{cc} for one puncture (based on \cite{na}) and ignore
the more extensive developments in \cite{cg,cb,co} based on 
\cite{dz,dg,dc,ka,kc,ko}.  We will only deal with the differential
$d\check{\Omega}_1$ as in (\ref{64})) plus the
canonical $d\tilde{\omega}$ (the curve will be taken in the form
$\eta^2=(\lambda-\Lambda^2)(\lambda+\Lambda^2)(\lambda-v)$ with
$d\tilde{v}=d\lambda/\eta$).  There will
be standard formulas for $F_{\gamma}$ with $d\tilde{S}
=T_1dS_{SW}
=T_1ad\tilde{\omega}+T_1d\check{\Omega}_1=\gamma d\tilde{\omega}+
T_1d\check{\Omega}_1$ as in (\ref{64}) ($d\check{\Omega}_1=
(1/\pi\sqrt{2})d\tilde
{\Omega}_1$) and $\gamma,\,T_1$ are independent variables with
$\partial d\tilde{S}/\partial\gamma=d\tilde{\omega}$ and $\partial
d\tilde{S}/\partial T_1=d\check{\Omega}_1$ while
\be
\gamma^D=F_{\gamma}^W=\frac{1}{2\pi i}\oint_Bd\tilde{S};\,\,
F^W_{\gamma\gamma}=\frac{1}{2\pi i}\oint_Bd\tilde{\omega}=\frac
{\tau}{2\pi i}
\label{100}
\ee
(factors of $2\pi i$ etc. come and go - these are consistent with
\cite{cc,na}, but
$F^{SW}_a=\oint_BdS_{SW}$ in \cite{ba} so
(\ref{49}) - (\ref{51}) may need adjustment when put into
the present framework - see remarks after (\ref{113})
below).  Further
\be
F^W_1=\partial_1F^W=-Res_{\infty}\,z^{-1}d\tilde{S};\,\,2F^W=\gamma
F_{\gamma}^W+T_1F^W_1=\frac{\tau}{2\pi i}
\gamma^2+2T_1\gamma F_{1\gamma}^W
+T_1^2F_{11}^W
\label{101}
\ee
and we recall from \cite{cc,na} that near $\lambda=\infty$ with
$z\sim\lambda^{-1/2}$ one can write
\be
d\tilde{\omega}=-\sum_1^{\infty}\sigma_{1,m}z^{m-1}da;\,\,d\check
{\Omega}_1=\left(-cz^{-2}-\sum_1^{\infty}q_{1,m}z^{m-1}\right)dz
\label{102}
\ee
(a multiplier $c=\sqrt{2}/\pi$ is needed here via (\ref{160}) below).
Next we note that 
\be
\partial_1F^W=-Res_{\infty}\,z^{-1}d\tilde{S}=
\label{103}
\ee
$$=Res_{\infty}z^{-1}
[\gamma\sum_1^{\infty}\sigma_{1,m}z^{m-1}+T_1(z^{-2}+
\sum_1^{\infty}q_{1,m}z^{m-1})]dz=\gamma\sigma_{1,1}+T_1q_{1,1}$$
and $\gamma \partial_{\gamma}F^W=(\gamma/2\pi i)\oint_Bd\tilde
{S}=\gamma^2\tau/2\pi i+(\gamma T_1/2\pi i)\oint_Bd\check{\Omega}_1$ while
the Riemann bilinear relations give
$(1/2\pi i)\oint_Bd\check{\Omega}_1=\sigma_{1,1}$ (cf. \cite{cc,na,sc}).
Consequently
\be
2F^W=\gamma F^W_{\gamma}+T_1F_1^W=\frac{1}{2\pi i}\gamma^2\tau+
2\gamma T_1\sigma_{1,1}+T_1^2q_{1,1}
\label{104}
\ee
Note also $(1/2\pi i)F_{\gamma 1}=(1/2\pi i)\oint_B\partial_1d\tilde{S}=
(1/2\pi i)\oint_Bd\check{\Omega}_1=\sigma_{1,1}$ so (\ref{104}) and
(\ref{100}) are compatible.
From the above we can also write
\be
2F^W=\gamma^2\frac{\partial^2F^W}{\partial\gamma^2}+2\gamma T_1\frac
{\partial^2F^W}{\partial\gamma\partial T_1}+T_1^2\partial_1^2F^W
\label{105}
\ee
and we emphasize (as in \cite{gf}) that $F^W$ is not $\underline{just}$
a quadratic function of $\gamma$ and $T_1$.  A nontrivial dependence
on these variables is expressed via the Whitham equations
(here e.g. ${\bf (L)}\,\,\partial_1d\tilde{\omega}=\partial_{\gamma}d
\check{\Omega}_1$) which involves the dependence of $d\tilde{\omega}$
and $d\check{\Omega}_1$ on $(\gamma,T_1)$ (recall also (\ref{33})
where the Whitham dynamics $(\clubsuit\clubsuit)$ of $u$ correspond to
the homogeneity condition for moduli).  In the present situation one
could write for $\Lambda$ constant
(modeled on (\ref{13}) and (\ref{64}))
\be
d\tilde{S}=T_1dS_{SW}=\gamma d\tilde{\omega}+T_1d\check{\Omega}_1\Rightarrow
\partial_1d\tilde{S}=dS_{SW}+T_1\frac{\partial dS_{SW}}{\partial v}
\frac{\partial v}{\partial T_1}=
\label{106}
\ee
$$=d\check{\Omega}_1\Rightarrow\oint_AdS_{SW}+T_1\frac{\partial v}
{\partial T_1}\oint_A\frac{\partial dS_{SW}}{\partial v}=0$$
But $\oint_A(\partial dS_{SW}/\partial v)=\partial a/\partial v
=\tilde{\sigma}$ from (\ref{67}) 
and consequently one obtains from the Whitham
dynamics of $v$
\be
\frac{\partial v}{\partial \,log(T_1)}=T_1\frac{\partial v}{\partial T_1}
=-\frac{a}{\tilde{\sigma}}
\label{107}
\ee
To use {\bf (L)} let us 
write (cf. \cite{cc,cv}))
\be
\partial_{\gamma}\left[z^{-2}+\sum_1^{\infty}q_{1m}z^{m-1}\right]
=\partial_1\sum_1^{\infty}\sigma_{1m}z^{m-1}\Rightarrow\partial_{\gamma}
q_{1m}=\partial_1\sigma_{1m}
\label{108}
\ee
and various arguments yield $q_{1m}=F^W_{1m}=\partial^2F^W/
\partial T_1\partial T_m$.  This can be based on 
asymptotics at $\infty$ using 
the Baker-Akhiezer (BA) function and $dKP$ formulas as in \cite{cc} 
(cf. also \cite{na}).  If $\partial_nS={\cal B}_n=\lambda^n_{+}$
with $\partial_1S=P$ corresponds to dKP 
(cf. \cite{ct}) then at $\infty$, $\partial_ndS
\sim d{\cal B}_n\sim\partial_nd\tilde{S}\sim d\check{\Omega}_n$.
Thus in particular the coefficients in (\ref{105}) depend on $\gamma$
and $T_1$ and in any event we can write $F^W$ as in (\ref{104}), expressed
entirely in terms of Riemann surface
(RS) quantities.  It is interesting that the natural
modulus $\tau$ appears here since this comes up in discussing the
beta function and many formulas are known (cf. \cite{bc,ba} and Section 3).
Note that formulas (\ref{41}) - (\ref{51}) for example (with $T_1=1$) go
into the present notation via $(y,v)\to (\eta,v)$.
\\[3mm]\indent
Now we have $F^W$ given entirely in terms of RS quantities via (\ref{104})
and this deserves attention.  Note also, if we regard the curve as arising
from a background 1-zone KdV situation one can also treat $-2F_{11}$ as the
dispersionless potential $U(T_1)$ arising from $u(x)$ in the Lax
operator $\partial^2-u(x)\,\,(x\to X=T_1$); we refer to \cite{ct,cs} for
discussion of this).  We remark further that the Whitham equations 
$\partial_{\gamma}d\check{\Omega}_1=\partial_1d\tilde{\omega}$
should lead to equations for branch point dependence on $(\gamma,T_1)$ as in
\cite{cc,fb}.  This means that $(\Lambda,v)$ are functions of $(\gamma,T_1)$
and the dependence of $\Lambda$ on $T_1$ has been neglected so far. 
In this direction we recall $d\check{\Omega}_1=(1/\pi\sqrt{2})(\lambda-
\tilde{\alpha})(d\lambda/\eta)$ where (cf. {\bf (EE)})
$a=(1/\pi\sqrt{2})\tilde{c}(\tilde{\alpha}-v)$ and 
$\tilde{c}d\tilde{\omega}=d\lambda/\eta$.  The Whitham equations
$\partial_1d\tilde{\omega}=\partial_{\gamma}d\check{\Omega}_1$ then
entail for $\Lambda=constant$
\be
\partial_1\frac{1}{\tilde{c}\eta}=\frac{1}{\pi\sqrt{2}}\partial_{\gamma}
\frac{\sqrt{\lambda-v}}{\sqrt{\lambda^2-\Lambda^4}}-\partial_{\gamma}
\frac{(a/\tilde{c})}{\eta}
\label{109}
\ee
After performing the derivations (with $\Lambda$ constant) one multiplies
by $(\lambda-v)^{3/2}$ and lets $\lambda\to v$; this yields
(using $\partial_a=T_1\partial_{\gamma}$ for $T_1$ constant)
\be
\partial_1v|_{\gamma,\Lambda}=-a\partial_{\gamma}v|_{1,\Lambda}\equiv 
T_1\partial_1v|_{\gamma,\Lambda}+
a\partial_av|_{1,\Lambda}=0
\label{110}
\ee
which is again the standard homogeneity condition for the modulus $v$.
On the other hand holding $v$ constant and repeating this procedure
one obtains, after multiplying by $(\lambda^2-\Lambda^4)^{3/2}$ and
letting $\lambda\to\Lambda^2$, we have
$
\partial_1\Lambda
=(\tilde{c}/\pi\sqrt{2})(\Lambda^2-v)\partial_
{\gamma}\Lambda-a\partial_{\gamma}\Lambda$
leading to
\be
T_1\partial_1\Lambda+a\partial_a\Lambda=
-2(\Lambda^2-v)\frac{\partial a}{\partial v}\frac{\partial\Lambda}
{\partial a}
\label{112}
\ee
Similarly, letting $\lambda\to-\Lambda^2$ one gets $T_1\partial_1\Lambda
+a\partial_a\Lambda=2(\Lambda^2+v)\partial_va\partial_a\Lambda$
so we conclude that 
$\partial_va\partial_a\Lambda\sim \partial_v\Lambda=0$ 
(as could be expected for $(v,\Lambda)$ independent)
and (\ref{112}) is again a homogeneity condition.
Thus, heuristically one arrives at
\\[3mm]\indent {\bf PROPOSITION 4.1.}$\,\,$  Under the hypotheses indicated,
we obtain from the Whitham equations $\partial_1d\tilde{\omega}=
\partial_{\gamma}d\check{\Omega}_1$, the equations
$T_1\partial_1v|_{\gamma,\Lambda}+a
\partial_av|_{1,\Lambda}=0$ for $\Lambda$ constant, 
and $T_1\partial_1\Lambda|_{\gamma,v}+a\partial_
a\Lambda|_{1,v}=0$ for $v$ constant.  These homogeneity equations
can also be obtained from the Whitham dynamics of $v$ as in (\ref{106})
or $\Lambda$ as in (\ref{118}) below.
\\[3mm]\indent
It would be nice now to work in $\tau$ in order to utilize the formulas
and techniques of \cite{bc,By,ba,Bz,be,ma}, some of which is indicated
in (\ref{41}) - (\ref{51}).  In this direction we note first from
(\ref{67}) and (\ref{106}) that ($d\lambda/\eta=\tilde{c}d\tilde{\omega}$)
$$
\frac{1}{2\pi i}\oint_Bd\check{\Omega}_1=\sigma_{11}=\frac{1}{2\pi i}
\oint_BdS_{SW}-\frac{T_1\partial_1v}{2\pi i 2\pi\sqrt{2}}\oint_B
\tilde{c}d\tilde{\omega}\Rightarrow$$
\be
\Rightarrow \sigma_{11}=a^D+T_1\tilde{\sigma}\partial_1v\tau=
a^D+T_1(\partial_1 a)\tau
\label{113}
\ee
Note again that in Section 3, $F_a^{SW}=\oint_BdS_{SW}$ but
in Section 4, $F_{\gamma}^W=(1/2\pi i)\oint_Bd\tilde{S}$ and we will
distinguish whenever necessary.  Now consider (in the notation of
Section 4)
\be
\partial_va^D=\frac{1}{2\pi i}\oint_B\frac{\partial dS_{SW}}
{\partial v}=-\frac{1}{2\pi i}\frac{1}{2\pi\sqrt{2}}\oint_B\frac{d\lambda}
{\eta}=\frac{\tau\tilde{\sigma}}{2\pi i}
\label{116}
\ee
Next we have in a straightforward manner
\be
\partial_v\sigma_{11}=\frac{\tau\tilde{\sigma}}{2\pi i}+
T_1(\partial_v\partial_1 a)\tau+T_1(\partial_1a)\partial_v\tau
=\left(\frac{\tilde{\sigma}}{2\pi i}+T_1\partial_1\tilde{\sigma}\right)\tau+
T_1(\partial_1a)\tau_v
\label{117}
\ee
Now for $v$ constant and variable $\Lambda$ (\ref{106}) corresponds to
\be
\partial_1d\tilde{S}=dS_{SW}+T_1\frac{\partial dS_{SW}}{\partial\Lambda}
\frac{\partial\Lambda}{\partial T_1}=d\check{\Omega}_1\Rightarrow 0=
\label{118}
\ee
$$=\oint_AdS_{SW}+T_1\frac{\partial\Lambda}{\partial T_1}\oint_A
\frac{\partial dS_{SW}}{\partial\Lambda}=a+T_1\partial_1\Lambda
\frac{\partial a}{\partial\Lambda}\sim a\frac
{\partial\Lambda}{\partial a}+T_1\partial_1\Lambda=0$$
which is again the homogeneity condition of Proposition 4.1, derived
now from the Whitham dynamics of $\Lambda$ (cf. (\ref{106})).
If we
multiply by $1/2\pi i$ and integrate over $B$ one gets further
\be
\sigma_{11}=a^D+T_1\frac{\partial\Lambda}{\partial T_1}
\partial_{\Lambda}a^D=a^D+T_1\partial_1a^D
\label{119}
\ee
(note this could also be concluded from (\ref{106}) as in (\ref{113})).  
\\[3mm]
\indent
Now one goal here is to relate $T_1$ and $\Lambda$ in the sense that
$T_1F_1^W=\Lambda F_{\Lambda}^W$.  Pragmatically this is what occurs
as shown in \cite{cb} (cf. also Remark 2.3), based on \cite{ba,ea,
ia,se} (cf. also \cite{ta}) and one wants also to 
reconcile this with formulas
like (\ref{32}) for example.
There is also another problem in determining what depends on what
and we will comment on this in Remark 5.7.
We have
seen in (\ref{5}) for example, or analogously in $\eta^2=(\lambda-v)
(\lambda^2-\Lambda^4)$ that many things can vary together.  
One sees that a priori $d\tilde{\omega}$
and $d\check{\Omega}_1$ depend only on $(v,\Lambda)$ while
$dS_{SW}=(\lambda-v)d\lambda/\pi\sqrt{2}$ depends only on $(\Lambda,v)$.
Hence $a=\oint_AdS_{SW},\,\,a^D=\oint_BdS_{SW},$ and
$\tau=\oint_Bd\tilde{\omega}$ depend only on $(\Lambda,v)$.  Now using
$\gamma$ and $T\sim T_1$ as Whitham times produces a dependence of 
$(\Lambda,v)$ on $(\gamma,T)$ so $\tau$ becomes a function of $(\gamma,
T)$ as do the coefficients $\sigma_{1m}$ and $q_{1m}$ in $d\tilde
{\omega}$ and $d\check{\Omega}_1$.  The defining equations for $a$
and perhaps $a^D$ should then be considered as ``constraints".  Note first
that no constraint arises from 
\be
\gamma=T_1a=\oint_AT_1dS_{SW}(T,\gamma)=\oint_Ad\tilde{S}(T,\gamma)=
\gamma\oint_Ad\tilde{\omega}=\gamma
\label{125}
\ee
which is tautological, while
\be
\gamma^D=\frac{1}{2\pi i}\oint_Bd\tilde{S}(T,\gamma)=\frac{\gamma}
{2\pi i}\oint_Bd\tilde{\omega}+\frac{T_1}{2\pi i}\oint_Bd\check
{\Omega}_1=\frac{\gamma\tau}{2\pi i}+T_1\sigma_{11}
\label{126}
\ee
which simply defines $\gamma^D$ as a function of $(\gamma,T_1)$.  Evidently
if $\partial\gamma^D/\partial T_1\not= 0$ one can 
determine $T_1=T_1(\gamma,
\gamma^D)$ and we record
\be
\partial_1\gamma^D=\frac{\gamma\tau_1}{2\pi i}+\sigma_{11}+T_1\partial_1
\sigma_{11}
\label{127}
\ee
(this is a reminder that it would be helpful to have
explicit formulas for $d\tilde{\omega},
\,\,\tau,$ and $d\check{\Omega}_1$).  To see the constraint aspect simply
look at the definition $a=\oint_AdS_{SW}(v,\Lambda)=\oint_A
dS_{SW}(a,T_1)$.  Explicitly one can look at (\ref{55}) to see that
\be
a=\sqrt{2}(v+\Lambda^2)^{1/2}F\left(-\frac{1}{2},\frac{1}{2},
1,\frac{2\Lambda^2}{v+\Lambda^2}\right)
\label{128}
\ee
with $(v,\Lambda)=(v,\Lambda)(\gamma,T_1)\sim (v,\Lambda)(a,T_1)$  
so $a$ is not
unrestricted (cf. Remark 5.7 for a few related comments).
Let us compute
now from (\ref{128}) to get 
\be
a_v=\frac{\sqrt{2}}{2}\frac{\sqrt{v+\Lambda^2}}{v+\Lambda^2}F+
\sqrt{2}\sqrt{v+\Lambda^2}F_4\left[\frac{-2\Lambda^2}
{(v+\Lambda^2)^2}\right];
\label{129}
\ee
$$a_{\Lambda}=\frac{\sqrt{2}\sqrt{v+\Lambda^2}2\Lambda F}
{2(v+\Lambda^2)}+\sqrt{2}\sqrt{v+\Lambda^2}F_4\left[\frac{4\Lambda}
{v+\Lambda^2}-\frac{4\Lambda^3}{(v+\Lambda^2)^2}\right]$$
Thus
\be
a_v=\frac{a}{2(v+\Lambda^2)}-\frac{2\sqrt{2}\Lambda^2
F_4}{(v+\Lambda^2)^{3/2}};\,\,a_{\Lambda}=\frac{\Lambda a}
{v+\Lambda^2}+\frac{\sqrt{2}F_44v\Lambda}{(v+\Lambda^2)^{3/2}}
\label{130}
\ee
leading to
\\[3mm]\indent {\bf PROPOSITION 4.2.}$\,\,$ The calculations above 
yield
\be
a_v-\frac{a}{2(v+\Lambda^2)}=-\frac{\Lambda}
{2v}\left[a_{\Lambda}-\frac{\Lambda a}{v+\Lambda^2}\right]
\Rightarrow
2va_v+\Lambda a_{\Lambda}=a
\label{131}
\ee
\indent
This seems quite charming and we observe that it is also a simple
consequence of $a(t^2v,t\Lambda)=ta(v,\Lambda)$.
In particular can we relate this to (\ref{67}) and (\ref{68}).
Thus 
\be
a_v=\tilde{\sigma}=
-\frac{1}{2\pi\sqrt{2}}\oint_A\frac{d\lambda}{\eta};\,\,
a=\frac{1}{\pi\sqrt{2}}\oint_A\frac{(\lambda-v)d\lambda}{\eta}
\label{132}
\ee
and we abandon here the formula (\ref{68}) for $a_{\Lambda}|_v$ since
it seems incompatible with (\ref{133}) below.
Note from (\ref{130}) the formula (to be used below)
$$
\tilde{\sigma}-\frac{a}{2(v+\Lambda^2)}=-\frac{2\sqrt{2}\Lambda^2
F_4}{(v+\Lambda^2)^{3/2}}\Rightarrow 
a_{\Lambda}=\frac
{\Lambda a}{v+\Lambda^2}-\frac{2v}{\Lambda}\left(\tilde{\sigma}-
\frac{a}{2(v+\Lambda^2)}\right)=$$
\be
=-\frac{2v\tilde{\sigma}}{\Lambda}+\frac{a}{v+\Lambda^2}\left(\Lambda
+\frac{v}{\Lambda}\right)=-\frac{2v\tilde{\sigma}}{\Lambda}+\frac{a}
{\Lambda}
\label{133}
\ee
\indent
{\bf REMARK 4.3.}$\,\,$ There seems to be some variety in notation
concerning hypergeometric representations of $a$ and
$a^D$.  Let us recall a few basic facts about hypergeometric
functions (cf. \cite{bd}).  Thus a standard notation is
\be
\int_0^1s^{b-1}(1-s)^{c-b-1}(1-zs)^{-a}ds=\frac{\Gamma(c-b)\Gamma(b)}
{\Gamma(c)}F(a,b,c,z)
\label{135}
\ee
For $a$ we will be dealing with $b-1=-1/2,\,\,c-b-1=-1/2,$ and $a=-1/2$ 
(thus $a=-1/2,\,\,b=1/2,$ and $c=1$) and we write (\ref{135}) via
$z=1/u$ and $s=(\zeta+\Lambda^2)/2\Lambda^2$ as
$$
\frac{1}{u^{1/2}}
\int_{-\Lambda^2}^{\Lambda^2}(\zeta+\Lambda^2)^{-1/2}(\Lambda^2-\zeta)^
{-1/2}\left(u-\frac{\zeta+\Lambda^2}{2\Lambda^2}\right)d\zeta=$$
\be
=\frac{\Gamma(1/2)\Gamma(1/2)}
{\Gamma(1)}F\left(-\frac{1}{2},
\frac{1}{2},1,\frac{2\Lambda^2}{v+\Lambda^2}\right)
\label{136}
\ee
Setting $u-(1/2)=v/2\Lambda^2$ and recalling
$\Gamma(1/2)=\sqrt{\pi}$ with $\Gamma(1)=1$ we get
\be
\pi F=\frac{1}{(2\Lambda^2u)^{1/2}}\int_{-\Lambda^2}^
{\Lambda^2}(\zeta+\Lambda^2)^{-1/2}(\zeta-\Lambda^2)^{-1/2}
(\zeta-v)^{1/2}d\zeta=
\label{137}
\ee
$$=\frac{1}{\sqrt{v+\Lambda^2}}\frac{\pi}{\sqrt{2}}
a(v,\Lambda)\Rightarrow a=\sqrt{2}(v+\Lambda^2)^{1/2}F$$
confirming (\ref{55}) for $a$ (cf. also \cite{kb,ra}).
\\[3mm]\indent {\bf REMARK 4.4.}$\,\,$  Let us also list here a few
more formulas from \cite{kb,ra} of possible use.  Thus in \cite{kb}
$\eta^2=(\lambda-v)(\lambda^2-\Lambda^4)$ and $\lambda_{SW}=(1/
2\pi\sqrt{2})(\lambda-v)^{1/2}(\lambda^2-\Lambda^4)^{-1/2}d\lambda$,
which is $(1/2)dS_{SW}$ in our sense, so we multiply the formulas of
\cite{kb} by $2$ or $1/2$ when appropriate.  Then
$$
a_v=\frac{1}{\sqrt{2}(v+\Lambda^2)^{1/2}}F\left(\frac{1}{2},
\frac{1}{2},1,\frac{2\Lambda^2}{v+\Lambda^2}\right)=
\frac{1}{\sqrt{2v}}F\left(\frac{1}{4},\frac{3}{4},1,\frac{\Lambda^4}
{v^2}\right);$$
\be
a^D_v=\frac{i}{\Lambda}F\left(\frac{1}{2},\frac{1}{2},1,\frac{\Lambda^2-v}
{2\Lambda^2}\right)
\label{138}
\ee
and for $a^D_v$ there is also an analytic continuation formula
given in \cite{kb} with argument $v^2/\Lambda^4$.
Another formula mentioned in \cite{kb} is
\be
\int_{e_1}^{e_2}\frac{dx}{[(x-e_1)(x-e_2)(x-e_3)]^{1/2}}=
\label{140}
\ee
$$=\frac{1}{\sqrt{e_3-e_1}}\int_0^1\frac{dt}{[t(1-t)(1-ht)]^{1/2}}
=\frac{\pi}{\sqrt{e_3-e_1}}F\left(\frac{1}{2},\frac{1}{2},1,h\right)$$
where $h=(e_2-e_1)/(e_3-e_1)$.  Another quantity of interest is
$\hat{\tau}=a^D/a$ which has the same monodromy as
$\tau=F^{SW}_{aa}$ (cf. \cite{ba}).  Evidently from
$a^D=F^{SW}_a$ one has also $F^{SW}_{aa}
=\tau=a^D_v/a_v$ (cf. \cite{kb} and note that 
in this section $a^D=(1/2\pi i)\oint_BdS_{SW}=F_a$
so $a^D/a=\tau/2\pi i$ as in (\ref{116})).
Further
formulas from \cite{ra} are (for $y^2=(x^2-1)(x-u)$ and $\lambda_{SW}
=(1/\pi\sqrt{2})(x-u)^{1/2}(x^2-1)^{-1/2}$)
\be
\check{a}=\sqrt{2}(u+1)^{1/2}F\left(-\frac{1}{2},
\frac{1}{2},1,\frac{2}{u+1}
\right);
\label{141}
\ee
$$\check{a}^D=\frac{i(u-1)}{2}F\left(\frac{1}{2},
\frac{1}{2},2,\frac{1-u}{2}\right)
=\frac{i(u-1)}{\sqrt{2}(u+1)^{1/2}}F\left(\frac{1}{2},\frac{3}{2},2,
\frac{u-1}{u+1}\right)$$
(one is interested in different regions of validity).  Further
\be
\check{a}_u=-\frac{1}{2\sqrt{2}\pi}F\left(\frac{1}{2},\frac{1}{2},1,
\frac{2}{u+1}\right);\,\,\check{a}^D_u=
\frac{i}{2}F\left(\frac{1}{2},\frac{1}{2},1,\frac{1-u}{2}\right)
\label{142}
\ee
and we can also write
\be
\check{a}_u=\frac{k}{\pi}K(k);\,\,\check{a}^D_u=\frac{ik}{\pi}K'(k);\,\,
\check{a}=\frac{4}{\pi k}E(k);\,\,\check{a}^D=
\frac{4i}{\pi k}[K'(k)-E'(k)]
\label{143}
\ee
where $k=[2/(u+1)]^{1/2}$ and $K,E$ are complete elliptic integrals of
the first and second kind.  The usual modular 
parameter $\tau_0=iK'/K$ becomes
$\check{a}^D_u/\check{a}_u$ which is the 
effective coupling $\tau=(\theta/2\pi)+(4i\pi/g^2)$
One checks easily now that (\ref{141}) gives the correct value
$a=\sqrt{2}\sqrt{v+\Lambda^2}F(-1/2,1/2,1,2\Lambda^2/(v+\Lambda^2))$
upon substitutions $u=v/\Lambda^2$ and $x=\lambda/\Lambda^2$.  For 
$\check{a}_u$ we must consider
$$
\check{a}_u=\frac{\partial}{\partial u}\oint_A\frac{(x-u)^
{1/2}dx}{\pi\sqrt{2}(x^2-1)^{1/2}}=
-\frac{1}{2\pi\sqrt{2}}\oint_A\frac{dx}{y}=$$
\be
=\frac{\Lambda}{\sqrt{2}(v+\Lambda^2)^{1/2}}F
\left(\frac{1}{2},\frac{1}{2},1,
\frac{2\Lambda^2}{v+\Lambda^2}\right)
\label{144}
\ee
Here from (\ref{132}) we note that $a_v=\tilde{\sigma}=
-(1/2\pi\sqrt{2})\oint_A(d\lambda/\eta)$ so
\be
\check{a}_u=-\frac{1}{2\pi\sqrt{2}}\oint_A\frac{dx}{y}=
-\frac{\Lambda}{2\pi
\sqrt{2}}\oint_A\frac{d\lambda}{(\lambda-v)^{1/2}
(\lambda^2-\Lambda^4)^{1/2}}
=\Lambda a_v
\label{145}
\ee
Hence we must have
\be
a_v=\frac{1}{\sqrt{2}(v+\Lambda^2)^{1/2}}F\left(\frac{1}{2},
\frac{1}{2},1,\frac{2\Lambda^2}{v+\Lambda^2}\right)=\tilde{\sigma}=
-\frac{\tilde{c}}{2\pi\sqrt{2}}
\label{146}
\ee
(in agreement with the first formula of (\ref{138})) and leading to
\be
\tilde{c}=-\frac{2\pi}{(v+\Lambda^2)^{1/2}}F\left(\frac{1}{2},\frac
{1}{2},1,\frac{2\Lambda^2}{v+\Lambda^2}\right)
\label{147}
\ee
in which form one must note that for the $A$ cycle $-\Lambda^2\leq\lambda
\leq\Lambda^2<v$ so
\be
\tilde{c}=\oint_A\frac{d\lambda}{\eta}=-\oint_A\frac{d\lambda}
{(v-\lambda)^{1/2}(\Lambda^4-\lambda^2)^{1/2}}
\label{148}
\ee
and we note for $v\geq\Lambda^2$ that $2\lambda^2/(v+\Lambda^2)\leq 1$ so the
argument of the hypergeometric function is reasonable.
\\[3mm]\indent
Next we want explicit formulas for $a_{\Lambda}$ and $\tau$ so 
consider first (using (\ref{140}) with $e_1=\Lambda^2,\,\,e_3=-\Lambda^2,$
and $e_2=v$)
\be
\tau=\oint_Bd\tilde{\omega}=\frac{1}{\tilde{c}}\oint_B\frac{d\lambda}
{\eta}\sim\frac{2}{\tilde{c}}\int_{\Lambda^2}^v\frac{d\lambda}
{(\lambda-v)^{1/2}(\lambda^2-\Lambda^2)^{1/2}}=
\label{149}
\ee
$$=\frac{2\pi}{\tilde{c}(-2\Lambda^2)^{1/2}}F\left(\frac{1}{2},\frac
{1}{2},1,\frac{v-\Lambda^2}{-2\Lambda^2}\right)=\frac{2\pi}{i\tilde{c}
\sqrt{2}\Lambda}F\left(\frac{1}{2},\frac{1}
{2},1,\frac{\Lambda^2-v}{2\Lambda^2}\right)$$
This should be compared then to $\tau=a^D_v/a_v$ or
(via (\ref{138}) and (\ref{146}))
\be
\tau=\frac{2\pi\sqrt{2}}{i\tilde{c}\Lambda}F\left(\frac{1}{2},\frac
{1}{2},1,\frac{\Lambda^2-v}{2\Lambda^2}\right)
\label{150}
\ee
Thus (\ref{149}) and (\ref{150}) agree if we multiply (\ref{149})
by $2$ (as indicated earlier for formulas of \cite{kb}).  Explicitly
now, using (\ref{147}) we have
\\[3mm]\indent {\bf PROPOSITION 4.5.}$\,\,$  Under the hypotheses
indicated
\be
\tau=\frac{i\sqrt{2}(v+\Lambda^2)^{1/2}}{\Lambda}\frac{F\left(
\frac{1}{2},\frac{1}{2},1,\frac{\Lambda^2-v}{2\Lambda^2}\right)}
{F\left(\frac{1}{2},\frac{1}{2},1,\frac{2\Lambda^2}{v+\Lambda^2}\right)}
\label{151}
\ee
The range of validity is $|(\Lambda^2-v)/2\Lambda^2|\leq 1$ 
and $|2\Lambda^2/(v+\Lambda^2)|\leq 1$ which gives a common region 
$\Lambda^2\leq v\leq 3\Lambda^2$.  For other values one would need 
analytic continuation formulas for example which exist in abundance.
\\[3mm]\indent
Next consider $a_{\Lambda}$ and we'll try to make sense of 
$(\ref{133})$.  First (\ref{133}) gives
(cf. (\ref{131}))
$$
a_{\Lambda}=-\frac{2v}{\Lambda}a_v+\frac{a}
{\Lambda}=-\frac{\sqrt{2}v}{\Lambda(v+\Lambda^2)^{1/2}}F\left(
\frac{1}{2},\frac{1}{2},1,\frac{2\Lambda^2}{v+\Lambda^2}\right)+$$
\be
+\frac{\sqrt{2}(v+\Lambda^2)^{1/2}}{\Lambda}F\left(-\frac{1}{2},
\frac{1}{2},1,\frac{2\Lambda^2}{v+\Lambda^2}\right)
\label{152}
\ee
Then
differentiate
$a$ in (\ref{128}) directly to get ($F=F(-1/2,1/2,1,2\Lambda^2/(v+
\Lambda^2))$)
\be
a_{\Lambda}=\sqrt{2}(v+\Lambda^2)^{-1/2}\Lambda F+
\frac{4\sqrt{2}\Lambda v}{(v+\Lambda^2)^{3/2}}F_4
\label{153}
\ee
Now for example (cf. \cite{bd})
\be
\frac{d}{dz}F(a,b,c,z)=\frac{ab}{c}F(a+1,b+1,c+1,z)
\label{154}
\ee
which implies
\be
a_{\Lambda}=\frac{\sqrt{2}\Lambda}{(v+\Lambda^2)^{1/2}}
F\left(-\frac{1}{2},\frac{1}{2},1,\frac{2\Lambda^2}{v+\Lambda^2}\right)
-\frac{\sqrt{2}v\Lambda}{(v+\Lambda^2)^{3/2}}F\left(\frac{1}{2},\frac
{1}{2},2,\frac{2\Lambda^2}{v+\Lambda^2}\right)
\label{155}
\ee
We note now the relation (cf. \cite{bd})
\be
cF(a,b,c,z)-cF(a+1,b,c,z)+bzF(a+1,b+1,c+1,z)=0\Rightarrow 
\label{156}
\ee
$$\Rightarrow F\left(-\frac
{1}{2},\frac{1}{2},1,z\right)-F\left(\frac{1}{2},\frac{1}{2},1,z\right)
+\frac{z}{2}F\left(\frac{1}{2},\frac{3}{2},2,z\right)=0$$
(where $z=2\Lambda^2/(v+\Lambda^2)$).  Using this in (\ref{155}) gives
$$
a_{\Lambda}=\frac{\sqrt{2}\Lambda}{(v+\Lambda^2)^{1/2}}F\left(
-\frac{1}{2},\frac{1}{2},1,\frac{2\Lambda^2}{v+\Lambda^2}\right)-\frac
{\sqrt{2}v}{\Lambda(v+\Lambda^2)^{1/2}}\times$$
\be
\times \left[F\left(\frac{1}{2},
\frac{1}{2},1,\frac{2\Lambda^2}{v+\Lambda^2}\right)-F\left(-\frac
{1}{2},\frac{1}{2},1,\frac{2\Lambda^2}{v+\Lambda^2}\right)\right]
\label{157}
\ee
Hence
\be
a_{\Lambda}=-\frac{v\sqrt{2}}{\Lambda(v+\Lambda^2)^{1/2}}F\left(\frac
{1}{2},\frac{1}{2},1,\frac{2\Lambda^2}{v+\Lambda^2}\right)+
\frac{\sqrt{2}(v+\Lambda^2)^{1/2}}{\Lambda}F\left(-\frac{1}{2},
\frac{1}{2},1,\frac{2\Lambda^2}{v+\Lambda^2}\right)
\label{158}
\ee
which agrees with (\ref{152}).
\\[3mm]\indent
We consider also now the dependence of the coefficients $\sigma_{11}$ and
$q_{11}$ on $(\Lambda,v)$ via $d\tilde{\omega}=-\sum_1^{\infty}
\sigma_{1m}z^{m-1}dz$ where $z\sim\lambda^{-1/2}$ and $d\check{\Omega}_1=
(-cz^{-2}-\sum_1^{\infty}q_{1m}z^{m-1})dz$.  
Here $d\check{\Omega}_1=(1/\pi\sqrt{2})
d\tilde{\Omega}_1=(1/\pi\sqrt{2})(\lambda-\tilde{\alpha})d\lambda/\eta$ and 
$\tilde{a}=(\tilde{c}/\pi\sqrt{2})(\tilde{\alpha}-v)$ (cf. {\bf (EE)}).
Thus for small $z\sim\lambda^{-1/2}$ one obtains 
(via $(1+x)^{-1/2}=1-(x/2)+
(3x^2/8)+\cdots$)
\be
\frac{d\lambda}{\eta}=\frac{d\lambda}{[(\lambda-v)
(\lambda^2-\Lambda^4)]^{1/2}}=
-2dz\left[1+\frac{1}{2}vz^2+\cdots\right]
\label{159}
\ee
Hence ${\bf (Q)}\,\,\sigma_{11}=2/\tilde{c}$.  Further we also have
\be
d\check{\Omega}_1=\frac{1}{\pi\sqrt{2}}(z^{-2}-\tilde{\alpha})
(-2-vz^2-\cdots)dz=\frac{dz}{\pi\sqrt{2}}[-2z^{-2}+(2\tilde{\alpha}-v)
+O(z)]
\label{160}
\ee
This gives the multiplier $c=\sqrt{2}/\pi$ of (\ref{102}) and exhibits
${\bf (R)}\,\,q_{11}=(v-2\tilde{\alpha})/\pi\sqrt{2}=-(v/\pi\sqrt{2})
-(2\tilde{a}/\tilde{c})=-(v/\pi\sqrt{2})-\tilde{a}\sigma_{11}$.
Summarizing gives
\\[3mm]\indent {\bf PROPOSITION 4.6.}$\,\,$  Under the hypotheses
indicated $q_{11}=-(v/\pi\sqrt{2})-\tilde{a}\sigma_{11}$ with
\be
\sigma_{11}=-\frac{(v+\Lambda^2)^{1/2}}{\pi F\left(\frac{1}{2},\frac{1}{2},
1,\frac{2\Lambda^2}{v+\Lambda^2}\right)};\,\,q_{11}=-\frac{v}{\pi\sqrt{2}}
+\frac{\sqrt{2}(v+\Lambda^2)}{\pi}\frac{F\left(-\frac{1}{2},\frac{1}{2},
1,\frac{2\Lambda^2}{v+\Lambda^2}\right)}{F\left(\frac{1}{2},\frac{1}{2},
1,\frac{2\Lambda}{v+\Lambda^2}\right)}
\label{161}
\ee
\indent
Finally, note that a formula for $\partial_{\Lambda}u$ arises in (\ref{29})
for $\delta a=0$ so in the present context we can write for $a$ fixed
$a_v|_{\Lambda}\delta v+a_{\Lambda}|_v\delta\Lambda =0$ where
$a_v|_{\Lambda}=\tilde{\sigma}$ is given in (\ref{146}) and 
$a_{\Lambda}|_v$ is given by (\ref{133}) or $(\ref{152})\equiv
(\ref{158})$.  From (\ref{133}) we get in particular (recall
$T_1\partial_1v=-(a/\tilde{\sigma})$ from (\ref{107}))
$a_{\Lambda}=-(2v\tilde{\sigma}/\Lambda)+(a/\Lambda)=
-(2v/\Lambda)a_v+(a/\Lambda)$ which implies
\\[3mm]\indent {\bf PROPOSITION 4.7.}$\,\,$  Under the hypotheses
indicated
\be
\tilde{\sigma}\frac{\partial v}{\partial\Lambda}+\frac{a}{\Lambda}-
\frac{2v\tilde{\sigma}}{\Lambda}=0\Rightarrow \Lambda v_{\Lambda}=
2v-\frac{a}{\tilde{\sigma}}=2v+T_1\partial_1v
\label{1611}
\ee
which corresponds exactly to (\ref{32}).

\section{$T_1$ AND $\Lambda$ DERIVATIVES OF $F$}
\renewcommand{\theequation}{5.\arabic{equation}}\setcounter{equation}{0}

We go now to $F^W$ as (\ref{104}) for example and will try to determine
explicitly the $T_1$ and $\Lambda$ derivatives in order to see 
how it is reasonable to write
$T_1\partial_1F^W=\Lambda\partial_{\Lambda}F^W$.
We doubt that $T_1\partial_1
F^W$ and $\Lambda\partial_{\Lambda}F^{SW}$ 
can be compared (cf. Remark 2.3)
so what must be clarified is how
to deal with the $T_1$ and $\Lambda$ dependence of quantities in $F^W$
and how to express the sense in which $\Lambda\sim T_1$.  
Thus
$2F^W=(1/2\pi i)T_1^2a^2\tau+2T_1^2a\sigma_{11}
+T_1^2q_{11}$ and \`a priori $a,\,\tau,\,\sigma_{11}$, and 
$q_{11}$ all depend on $T_1$ through Whitham dynamics.  Thus from
(\ref{110}) and Proposition 4.1 one has $T_1\partial_1v+a\partial_a
v=0$ with $T_1\partial_1\Lambda+a\partial_a
\Lambda=0$.  Here the relation $\partial_a=T_1\partial_{\gamma}$ has
been used which is appropriate since we deal with 
$\partial_{\gamma}$ for $T_1=c$ and $\partial_1$ for $\gamma=c$.
Consider now $\sigma_{11}=2/\tilde{c}=-1/\pi\sqrt{2}\tilde{\sigma}$ and
$q_{11}=-(v/\pi\sqrt{2})-a\sigma_{11}=-(v/\pi\sqrt{2})+
(a/\pi\sqrt{2}\tilde{\sigma})$ so one has
\be
2F^W=\frac{T_1^2}{2\pi i}a^2\tau+2T_1^2a\sigma_{11}
+T_1^2\left(-\frac{v}{\pi\sqrt{2}}-a\sigma_{11}\right)=
\label{163}
\ee
$$=-\frac{T_1^2v}{\pi\sqrt{2}}+\frac{\gamma^2\tau}{2\pi i}-
\frac{T_1\gamma}{\pi\sqrt{2}\tilde{\sigma}}$$
We can assume $\gamma$ and $T_1$ are independent to obtain
\be
\partial_1F^W=-\frac{2T_1v}{\pi\sqrt{2}}-\frac{T_1^2\partial_1v}{\pi\sqrt{2}}
+\frac{\gamma^2\partial_1\tau}{2\pi i}-\frac{\gamma}{\pi\sqrt{2}\tilde
{\sigma}}-\frac{T_1\gamma}{\pi\sqrt{2}}\partial_1\tilde{\sigma}^{-1}
\label{164}
\ee
On the other hand $F^W(T_1=1)\sim F^{SW}$ with $\gamma\to a$ and
e.g.
\be
\left.\partial_{\Lambda}F^{SW}\right|_v=\frac{1}{2\pi i}\left.\partial_
{\Lambda}(a^2\tau)\right|_v-\left.\partial_{\Lambda}
\left(\frac{a}{\pi\sqrt{2}\tilde{\sigma}}\right)\right|_v
\label{165}
\ee
For the $\partial_1$ derivatives in (\ref{164}) one can in principle
assume $\gamma$ fixed with $(v,\Lambda)$ functions of $T_1$.  Hence

\be
\partial_1F^W=-\frac{\sqrt{2}T_1v}{\pi}-\frac{T_1^2\partial_1v}{\pi
\sqrt{2}}-\frac{\gamma}{\pi\sqrt{2}\tilde{\sigma}}+
\label{166}
\ee
$$+\frac{\gamma^2}{2\pi i}\left[\tau_v|_{\Lambda}v_1+\tau_{\Lambda}|_v
\Lambda_1\right]-\frac{T_1\gamma}{\pi\sqrt{2}}\left[\partial_v
\tilde{\sigma}^{-1}|_{\Lambda}v_1+\partial_{\Lambda}\tilde{\sigma}^{-1}|_v
\Lambda_1\right]$$
We have at our disposal now explicit formulas for the $v$ and $\Lambda$
derivatives as well as relations such as $4(\Lambda^4-v^2)\tilde{\sigma}_v
=a,\,\,T_1v_1=-a/\tilde{\sigma}$, etc.
First consider $\partial_{\Lambda}F^{SW}$ from (\ref{165}) in the form
($v$ constant)
\be
\partial_{\Lambda}F^{SW}=\frac{1}{2\pi i}\left[2aa_{\Lambda}
\tau+a^2\tau_{\Lambda}\right]-\frac{1}{\pi\sqrt{2}}\left(\frac
{a_{\Lambda}}{\tilde{\sigma}}-\frac{a\tilde{\sigma}_
{\Lambda}}{\tilde{\sigma}^2}\right)
\label{167}
\ee
From {\bf (T)} we have now $\Lambda a_{\Lambda}=a-
2v\tilde{\sigma}$ while $\Lambda\tau_{\Lambda}=\beta(\tau)$ from
(\ref{44}).  For $\tilde{\sigma}=-\tilde{c}/2\pi\sqrt{2}$
we consider (cf. (\ref{147}))
\be
\tilde{c}_{\Lambda}=-\frac{\Lambda\tilde{c}}{v+\Lambda^2}=\frac{4\pi
\Lambda vF_4}{(v+\Lambda^2)^{5/2}}
\label{168}
\ee
This $F_4$ is not the same as that arising in (\ref{128}) - (\ref{129})
but since we know $\tilde{\sigma}_v$ we can eliminate $F_4$ via
$\tilde{c}_v$.  Thus
\be
\tilde{c}_v=-\frac{\tilde{c}}{2(v+\Lambda^2)}+\frac{4\pi
\Lambda^2F_4}{(v+\Lambda^2)^{5/2}}
\label{169}
\ee
This implies (after some calculation)
\be
\tilde{c}_{\Lambda}=-\frac{v}{\Lambda}\tilde{c}_v
-\frac{\tilde{c}}{2\Lambda}
\left[1+\frac{\Lambda^2}{v+\Lambda^2}\right]
\label{170}
\ee
\indent {\bf PROPOSITION 5.1.}$\,\,$  We can write (\ref{170})
in two ways.  First directly
\be
\Lambda\tilde{c}_{\Lambda}+v\tilde{c}_v=-\frac{\tilde{c}}{2}
\left[1+\frac{\Lambda^2}{v+\Lambda^2}\right]
\label{171}
\ee
and second, using $\tilde{c}=-2\pi\sqrt{2}\tilde{\sigma}$ and 
$4(\Lambda^4-v^2)\tilde{\sigma}_v=a$ one has
\be
\Lambda\tilde{c}_{\Lambda}+\frac{\tilde{c}}{2}\left[1+\frac{\Lambda^2}
{v+\Lambda^2}\right]=-v\tilde{c}_v=\frac{\pi va}
{\sqrt{2}(\Lambda^4-v^2)}
\label{172}
\ee
\indent
This now gives us all the ingredients needed for (\ref{167}).  Thus,
writing $\tilde{\sigma}_{\Lambda}=-\tilde{c}_{\Lambda}/2\pi\sqrt{2}$, we
have
\be
\Lambda\partial_{\Lambda}F^{SW}=\frac{1}{2\pi i}\left[2a\tau
(a-2v\tilde{\sigma})+a^2\beta(\tau)\right]-
\label{173}
\ee
$$-\frac{1}{\pi\sqrt{2}}\left\{\frac{a-2v\tilde{\sigma}}
{\tilde{\sigma}}+\frac{a}{2\pi\sqrt{2}\tilde{\sigma}^2}\left[
\frac{\pi va}{\sqrt{2}(\Lambda^4-v^2)}+\pi\sqrt{2}\tilde{\sigma}
\left(1+\frac{\Lambda^2}{v+\Lambda^2}\right)\right]\right\}$$
which is of possible interest due to the presence of the beta
function.
Consequently we state
\\[3mm]\indent {\bf PROPOSITION 5.2.}$\,\,$  From (\ref{173}) follows
\be
\Lambda\partial_{\Lambda}F^{SW}=\frac{1}{2\pi i}\left[2a^2\tau
-4va\tilde{\sigma}\tau +a^2\beta(\tau)\right]-
\label{174}
\ee
$$-\frac{1}{\pi\sqrt{2}}\left[\frac{a}{\tilde{\sigma}}-2v
+\frac{va^2}{4\tilde{\sigma}^2(\Lambda^4-v^2)}+\frac{a}
{\tilde{\sigma}}\left(1+\frac{\Lambda^2}{v+\Lambda^2}\right)\right]$$
\\[3mm]\indent
Now we want to check the vague formula ${\bf (U)}\,\,\Lambda\partial_
{\Lambda}F=-(2iu/\pi)=2F-aF_a|_{\Lambda}$ which is 
supposed to agree somehow with $T_1\partial_1F$ (cf. \cite{ba,Bz,cb,
co,ea,gf,ia,ma,na,se,ta}).  It is not clear here \`a priori how this
role $T_1\sim \Lambda$ arises or what the exact statement should be.
In particular {\bf (U)} is presumably valid 
for $F=F^{SW}$ but $\partial_1F$ is
defined only for $F=F^W$.  Moreover $F^{SW}\sim F^W(T_1\to 1)$ so it
is not clear how $\Lambda\sim T_1$ can be managed. 
First let us clarify some matters of homogeneity. 
The argument is made
in \cite{ba,se} that since $\tau=F_{aa}$ is dimensionless one has $aF_a|_
{\Lambda}+\Lambda F_{\Lambda}|_a=2F$ (where $F\sim F^{SW}$). 
In dealing with $F^{SW}\sim F$ as a function of 
$(a,\Lambda)$ we are using $F=F(v,\Lambda)$ with $a=a(v,\Lambda)$.  Then
if $a_v\not= 0$ one solves for $v=v(a,\Lambda)$ and puts this in
$F=F(v,\Lambda)\sim F(a,\Lambda)$.  Recall also $a_v\sim
\tilde{\sigma}$ so generically this should be satisfactory.  
As noted after Proposition
4.2, $a(t^2v,t\Lambda)=ta(v,\Lambda)$, leading to
$2va_v+\Lambda a_{\Lambda}=a$ and similarly
one checks that $\tilde{\sigma}(t^2v,t\Lambda)=(1/t)\tilde{\sigma}
(v,\Lambda)$ which implies $2v\tilde{\sigma}_v+\Lambda\tilde{\sigma}_
{\Lambda}=-\tilde{\sigma}$.  Since $\tilde{\sigma}=a_v=
-(1/2\pi\sqrt{2})\oint_A(d\lambda/\eta)$ and $\tau\tilde{\sigma}=
-(1/2\pi i)(1/2\pi\sqrt{2})\oint_B(d\lambda/\eta)$ (cf. (\ref{116}))
we have also $2v(\tau\tilde{\sigma})_v+\Lambda(\tau\tilde{\sigma})_
{\Lambda}=-\tau\tilde{\sigma}$.  Consequently
\\[3mm]\indent {\bf PROPOSITION 5.3.}$\,\,$  From the formulas indicated
follows the homogeneity $2v\tau_v+\Lambda\tau_{\Lambda}=0$ and
from $a(t^2v,t\Lambda)=ta(v,\Lambda)$ one gets
$\tau(ta,t\Lambda)\sim\tau(t^2v,t\Lambda)=\tau(v,\Lambda)
\sim\tau(a,\Lambda)$ which means $a\tau_a+
\Lambda\tau_{\Lambda}=0$ (the dimensionless property).
\\[3mm]\indent
One sees also that homogeneity properties can be determined from the
hypergeometric representations directly.  For example in (\ref{151}),
setting $\Lambda=t\Lambda_0$ and $v=t^2v_0$ the argument in the 
hypergeometric functions is unchanged as in the multiplier in front.
Hence $\tau(t^2v,t\Lambda)=\tau(v,\Lambda)$.  Further from (\ref{161})
$\sigma_{11}(t^2v,t\Lambda)=t\sigma_{11}(v,\Lambda)$ so
${\bf (V)}\,\,2v\partial_v\sigma_{11}+\Lambda\partial_{\Lambda}
\sigma_{11}=\sigma_{11}$ while $q_{11}=-(v/\pi\sqrt{2})-a
\sigma_{11}$ satisfies $q_{11}(t^2v,t\Lambda)=t^2q_{11}(v,\Lambda)$
(which is also obvious from (\ref{161})) leading to
${\bf (W)}\,\,2v\partial_vq_{11}+\Lambda\partial_{\Lambda}q_{11}=
2q_{11}$.  As a result one can look at $2F^W =T^2_1[(a
\tau/2\pi i)+2a\sigma_{11}+q_{11}]$ to see
that $F^W(t^2v,t\Lambda)=t^2F^W(v,\Lambda)$ yielding
\\[3mm]\indent {\bf PROPOSITION 5.4.}$\,\,$  For fixed $T_1$ the
prepotential $F^W$ satisfies $2v\partial_vF^W+\Lambda\partial_{\Lambda}
F^W=2F^W$.
\\[3mm]\indent
Further since $(v,\Lambda)\to v(a,\Lambda)$
with $a(t^2v,t\Lambda)=ta(v,\Lambda)$ we can write
for $T_1$ fixed $F^W(v,\Lambda)=F^W(a,\Lambda)$ (in an
obvious abuse of notation).  It follows that $F^W(ta,t\Lambda)\sim
F^W(t^2v,t\Lambda)=t^2F^W(v,\Lambda)\sim t^2F^W(a,\Lambda)$.
Consequently
\\[3mm]\indent
{\bf COROLLARY 5.5.}$\,\,$  For fixed $T_1,\,\,F^W$ satisfies
$aF^W_a+\Lambda F^W_{\Lambda}=2F^W$.
\\[3mm]\indent
One notes that if $F(a,\Lambda)$ is homogeneous of degree two then 
$F_{aa}$ is homogeneous of degree zero.  The converse would produce
$F$ of degree $2$ from $\tau\sim F_{aa}$ of degree $0$ and for suitable
$F$ this follows frorm $(1/t)f(ta)=\int_0^af_{a'}(ta')da'=
\int_0^af_{a'}(a')da'=f(a)$.
Now the general Whitham theory produces $T_1F_1^W+\gamma F^W_{\gamma}
=2F^W$ (cf. \cite{cc,ia,na}) and for $T_1$ fixed with $\gamma=
T_1a,\,\,F_a=T_1F_{\gamma}$ yielding ${\bf (\Phi)}\,\,
T_1F_1^W+aF^W_a=2F^W$; consequently
\\[3mm]\indent {\bf PROPOSITION 5.6.}$\,\,$  The 
identification $T_1F_1^W\sim
\Lambda F^W_{\Lambda}$ follows from ${\bf (\Phi)}$ and Corollary 5.5.
\\[3mm]\indent
{\bf REMARK 5.7.}$\,\,$  There are still 
various questions of what depends on what
and we consider a few aspects of this here.  From Proposition 4.1
we know $T_1\partial_1v|_{\Lambda,\gamma}+a\partial_av|_{\Lambda,1}=0$
and $T_1\partial_1\Lambda|_{\gamma,v}+a\partial_a\Lambda|_{1,v}=0$
while from (\ref{128}) for example one has a ``constraint" 
$a=a(v,\Lambda)$
and $(v,\Lambda)=(v,\Lambda)(\gamma,T_1)\sim (v,\Lambda)(a,T_1)$.
Now first, for say $v(\gamma,T_1)$, one can
write $\delta v=v_{\gamma}|_1\delta\gamma+v_1|_{\gamma}\delta T_1$ and
for $T_1=c,\,\,v_{\gamma}=(1/T_1)v_a$ with $\delta\gamma=T_1\delta a$
so $\delta v=v_a|_1\delta a+v_1|_{\gamma}\delta T_1$.  Similarly
$\delta\Lambda=\Lambda_a|_1\delta a+\Lambda_1|_{\gamma}\delta T_1$ and 
the Jacobian of the map $(a,T_1)\to (v,\Lambda)$ apparently is, by
homogeneity, $\Delta=v_1\Lambda_a-v_a\Lambda_1=0$,
which seems curious.  Indeed it is wrong since the homogeneity relations
explicitly require $\Lambda$ or $v$ fixed so in fact ${\bf (\Delta)}
\,\,\Delta=v_1|_{\gamma}\Lambda_a|_1-v_a|_1\Lambda_1|_{\gamma}$ and 
$v_a|_1$ for example refers to $\partial_av(a,T_1)|_1$ which is
a priori quite different from $v_a|_{1,\Lambda}$ in the homogeneity
condition.  Secondly, from $\gamma=T_1a(v,\Lambda)=k$ there results
\be
a+T_1a_v|_{\Lambda,\gamma}v_1|_{\Lambda,\gamma}+T_1a_{\Lambda}|_{\gamma,v}
\Lambda_1|_{\gamma,v}=0
\label{176}
\ee
This relation was very productive when either $\Lambda$ or $v$ is held
constant, as indicated in (\ref{107}) and (\ref{118}).  We note that
$v_a|_{1,\Lambda}$ is not the same as $v_a|_{\gamma,\Lambda}$ so the
homogeneity conditions only produce a formula
${\bf (Y)}\,\,a_v|_{\gamma,\Lambda}v_a|_{1,\Lambda}+a_{\Lambda}|_
{\gamma,v}\Lambda_a|_{1,v}=1$, of no apparent use.

\end{document}